\documentclass[journal]{IEEEtran}

\usepackage{graphicx, graphics, times, amsfonts, amsmath, amssymb, cite}
\usepackage{amsthm}
\usepackage{color}
\usepackage{hyperref}
\usepackage{multirow}
\usepackage{rotating}
\usepackage[capitalise]{cleveref}
\usepackage{subcaption}
\usepackage{bm}
\input{mysymbol.sty}
\usepackage[utf8]{inputenc}
\usepackage{enumitem}
\usepackage[ruled,linesnumbered]{algorithm2e}

\usepackage{moresize}

\usepackage{tikz}
\usetikzlibrary{cd}
\usetikzlibrary{shapes.geometric}
\usetikzlibrary{arrows.meta}
\usepgflibrary{plotmarks}
\usetikzlibrary{positioning}
\usetikzlibrary{backgrounds}
\usetikzlibrary{matrix}
\usetikzlibrary{decorations.markings}
\usetikzlibrary{calc}

\usepackage{pgfplots}
\pgfplotsset{compat=1.17}
\pgfplotstableset{col sep=comma}
\usepgfplotslibrary{groupplots}
\usepackage{pdfpages}

\newtheorem{theorem}{Theorem}

\newtheorem{proposition}{Proposition}

\newtheorem{problem}{Problem}

\newcommand\EE{\mathbb{E}}

\begin{document}

\title{Enhanced graph-learning schemes driven by similar distributions of motifs}

\author{{Samuel Rey,~\IEEEmembership{Student Member,~IEEE},
        T. Mitchell Roddenberry,~\IEEEmembership{Student Member,~IEEE}, Santiago Segarra,~\IEEEmembership{Member,~IEEE}, 
        and~Antonio G. Marques,~\IEEEmembership{Senior Member,~IEEE}}
\thanks{Work in this paper was partially supported by the Spanish Grants SPGRAPH (PID2019-105032GB-I00), FPU17/04520 and EST21/00420, and the USA NSF award CCF-2008555.}}

\maketitle

\begin{abstract}
This paper looks at the task of network topology inference, where the goal is to learn an unknown graph from nodal observations. One of the novelties of the approach put forth is the consideration of \emph{prior information} about the \emph{density of motifs} of the unknown graph to enhance the inference of classical Gaussian graphical models.
Dealing with the density of motifs directly constitutes a challenging combinatorial task.
However, we note that if two graphs have similar motif densities, one can show that the expected value of a polynomial applied to their empirical spectral distributions will be similar. Guided by this, we first assume that we have a reference graph that is related to the sought graph (in the sense of having similar motif densities) and then, we exploit this relation by incorporating a similarity constraint and a regularization term in the network topology inference optimization problem. The (non-)convexity of the optimization problem is discussed and a computational efficient alternating majorization-minimization algorithm is designed. We assess the performance of the proposed method through exhaustive numerical experiments where different constraints are considered and compared against popular baselines algorithms on both synthetic and real-world datasets. 
\end{abstract}


\tikzset{every mark/.append style={scale=1.5, solid}, font=\footnotesize}

\pgfplotsset{
    width=1.05\textwidth,
    legend style={
        font=\ssmall ,  
        inner xsep=1pt,
        inner ysep=1pt,
        nodes={inner sep=1pt}},
    legend cell align=left,
	every axis/.append style={line width=0.5pt},
	every axis plot/.append style={line width=1.25pt},
    every axis y label/.append style={yshift=-6pt}
}

\section{Introduction}\label{S:intro}
\IEEEPARstart{H}{arnessing} graphs to model the underlying structure of signals is gaining relevance due to the rising of data defined over non-Euclidean domains.
This graph-based perspective is at the heart of machine learning over graphs and graph signal processing (GSP), fields devoted to the development of tools for processing and learning from signals defined over irregular supports modeled by graphs~\cite{shuman2013emerging,sandryhaila2013discrete,ortega2018graph,djuric2018cooperative}.
Successful applications of these tools are found in power, communication, social, geographical, financial, and brain networks, to name a few~\cite{kolaczyk2009book,shuman2013emerging,sporns2012book,nodop1998field}.
While the default approach is to assume that the graph is known and focus on the processing of the network data, there are many relevant scenarios where the topology of the graph is unknown. To handle this, a preliminary (critical) step is to learn the topology of the graph from a set of nodal observations.  
The key to this task, which is commonly known as \emph{network topology inference} or \emph{graph learning}, is to leverage models/assumptions relating the properties of the observed signals to the topology of the sought graph~\cite{segarra2017network,segarra2018network,mateos2019connecting,sardellitti2019graph,buciulea2019network}.
Noteworthy approaches to this task include partial correlations and Gaussian graphical models~\cite{lauritzen1996graphical,friedman2008sparse,danaher2014joint,egilmez2017graph,kumar2019structured}, sparse structural equation models~\cite{cai2013sparse,baingana2014proximal}, smooth (total variation) 
models~\cite{kalofolias2016learn,dong2016learning,saboksayr2021accelerated}, and graph stationary models~\cite{segarra2017network,shafipour2020online,roddenberry2021network,buciulea2021learning}, among others.

All the aforementioned graph-learning approaches share one common characteristic: the focus is placed on the \emph{signals} rather than the graphs.
Indeed, most works learn the graph that best explains the observations without considering any prior information about the topology of the graph other than its sparsity.
If information about the topological structure of the graph is available, we can harness it to improve the quality of the estimated graphs by promoting desired structural characteristics.
An initial step in this direction is taken in joint graph-learning algorithms~\cite{danaher2014joint,navarro2020joint,yang2020network,rey2021joint}, where several graphs are jointly estimated under the additional assumption that they are close to each other in some sense.
This assumption is indeed justified when, e.g., the graphs being estimated proceed from the same distribution.
Nonetheless, measuring the distance between two graphs is a non-trivial endeavor and joint inference works are typically constrained to comparing graphs with a common set of nodes and promoting similar edge support across all graphs.

Some other works are also starting to take into consideration prior information about the graph.
A relevant example is found in \cite{kumar2019structured}, where the authors propose recovering the graph Laplacian from a set of Gaussian Markov random field (GMRF) observations while \emph{promoting desired properties over its spectrum}.
However, the convex constraints employed in the paper are limited to capturing basic information about the spectrum such as the number of zero eigenvalues.
Later on, \cite{roddenberry2021network} introduces a different graph learning method where the unknown graph is assumed to be drawn from a graphon.
The main limitations of such an approach are that the graphon is assumed to be known, which may not be trivial in practice since it involves knowing the distribution of the unknown graph, and moreover, that not every graph may be represented as a graphon.

To overcome previous limitations, in this paper we propose a novel graph-learning algorithm that considers prior information about the topology of the graph in a general yet informative way.
We start with the assumption that a reference graph with a \emph{density of motifs similar} to that of the sought graph is known, and we harness it to reveal a connection between the spectra of both graphs.
This allows us to circumvent the challenges associated with the combinatorial nature of the density of motifs.
Then, we approach the graph learning task by means of an optimization problem where we exploit the spectral similarity of the reference and the sought graph as a constraint.
Because the resulting algorithm is derived from the density of motifs it is local in nature, which allows us to compare graphs of different sizes (as described in further detail in later sections).
Furthermore, the proposed similarity constraints involve the \emph{distribution} of the eigenvalues, which results in constraints that are more informative than the ones considered in previous works.

After reviewing basic ideas in graph signal processing and graph learning in \cref{S:preliminaries}, the structure and main contributions of the paper are summarized next:
\begin{enumerate}
    \item We relate the structural characteristics of a graph described by the density of motifs to the graph spectrum (\cref{S:density_to_spectrum}).
    \item With this relationship in mind, we propose an optimization program for network topology inference (\cref{S:convex_relaxation}).
    \item Due to the nonconvexity of the problem, we specify an alternating Majorization-Minimization algorithm (\cref{S:algorithm}).
\end{enumerate}
Interesting generalizations of the considered graph learning problem are discussed in \cref{S:BeyondGMRF}, and then, the effectiveness of the proposed approach is demonstrated in \cref{S:experiments}, followed by brief concluding remarks.

\section{Notation and preliminaries: Graphs, GSP and GMRFs}\label{S:preliminaries}

We briefly introduce graph and GSP-related notation and review the definition of GMRFs.  

\vspace{1mm}
\noindent\textbf{Graphs}: 
Let $\ccalG:=(\ccalV,\ccalE)$ denote an undirected and weighted graph with a set of nodes $\ccalV$ and a set of edges $\ccalE$.
The graph is composed of $|\ccalV|=N$ nodes and, for every $i,j\in\ccalV$, we have that $(i,j)\in\ccalE$ if and only if the nodes $i$ and $j$ are directly connected.
The neighborhood of any node $i$ represents the set of nodes that are connected to $i$, i.e., $\ccalN_i := \{j \in \ccalV| (i,j) \in \ccalE \}$.
The connectivity of $\ccalG$ is captured in the sparse adjacency matrix $\bbA\in\reals^{N \times N}$ with $A_{ij} = 0$ only if $(i,j)\not\in\ccalE$, and whose entry $A_{ij}$ represents the weight of the edge between nodes $i$ and $j$.

\vspace{1mm}
\noindent\textbf{Graph signals and GSP}: 
Together with the graph $\ccalG$, we consider signals defined on (associated with) $\ccalV$, the nodes of $\ccalG$. Formally, a \emph{graph signal} can be modeled as a function from the vertex set to the real field $x:\ccalV\to\reals$ or, equivalently, as an $N$-dimensional vector $\bbx\in\reals^N$, with $x_i$ denoting the signal value at node $i$.
The last key element in the GSP framework is the so-called \emph{graph-shift operator} (GSO), an $N \times N$ matrix denoted as $\bbS$~\cite{sandryhaila2013discrete}.
The GSO, whose entries satisfy that $S_{ij}$ can be non-zero only if $i=j$ or $(i,j)\in\ccalE$, captures the topology of the underlying graph $\ccalG$ and can be understood as a topology-aware local operator that can be applied to process graph signals. Typical choices for the GSO include the adjacency matrix $\bbA$, the graph combinatorial Laplacian $\bbL:=\diag(\bbA\mathbf{1}) - \bbA$, and its normalized variants~\cite{shuman2013emerging,sandryhaila2013discrete}.
Note that $\diag(\cdot)$ denotes the diagonal operator that transforms a vector into a diagonal matrix and $\mathbf{1}$ denotes the vector of all ones.
Since $\ccalG$ is undirected, it follows that $\bbS$ is symmetric and it can be diagonalized as $\bbS=\bbV\bbLambda\bbV^\top$, where the orthonormal matrix $\bbV\in\reals^{N \times N}$ collects the eigenvectors of $\bbS$, and the diagonal matrix $\bbLambda=\diag(\bblambda)$ collects the eigenvalues $\bblambda\in\reals^N$.

\vspace{1mm}
\noindent\textbf{GMRF}:
A multivariate normal distribution is said to form a GMRF with respect to a graph $\ccalG=(\ccalV,\ccalE)$ if the edges not present in $\ccalE$ correspond to zeros on the precision matrix (the inverse covariance matrix). Upon selecting the GSO $\bbS$ as the positive definite precision matrix, the previous definition implies that if the random graph signal $\bbx$ follows a multivariate normal distribution $\ccalN(\mathbf{0},\bbS^{-1})$, then $\bbx$ is a GMRF with respect to $\bbS$.  

As a result, the probability density function (PDF) of a zero-mean GMRF with GSO $\bbS$ is simply
\begin{equation}\label{E:gaussian_pdf_S} 
f_{\bbx}(\bbx; \bbS) = (2\pi)^{-N/2} \det(\bbS)^{1/2} \exp\left(-\frac 1 2 \bbx^\mathrm{T}{\bbS}\bbx\right).
\end{equation}
The above expression will be critical to postulate an optimization that learns (estimates) the GSO $\bbS$ (and, hence, the edge set $\ccalE$) from nodal observations, a key question at the core of Gaussian graphical models~\cite{lauritzen1996graphical,friedman2008sparse,ravikumar2011high}.

\section{Graph learning from motif similarity}\label{S:gl_motifs}

Suppose now that we have access to a collection of $M$ graph signals $\bbX=[\bbx_1,...,\bbx_M]$. Each of the $M$  signals collects $N$ measurements (one per node) associated with the nodes of a graph $\ccalG$ that is not known. The graph learning problem aims at using $\bbX\in\reals^{N\times M}$ to estimate the GSO $\bbS\in\reals^{N \times N}$ and, as a result, to identify the \emph{unknown} edge set $\ccalE$ that connects the nodes in the graph $\ccalG$. To render this problem tractable, we consider two main assumptions:

\begin{itemize} 
\item The first one is that we have prior information on the (local properties) of the graph $\ccalG$ and, in particular, on the distribution of its motifs. More especifically, we consider having access to some reference graph $\tilde{\ccalG}$ with a \emph{density of motifs similar} to that of the unknown graph $\ccalG$. Understanding a graph as a composition of motifs arouses a particular interest due to the \emph{local} nature of motifs~\cite{roddenberry2022local}. Intuitively, assuming that two graphs have a similar density of motifs can be interpreted as assuming that both graphs have common ``building blocks'' or similar patterns. 
\item The second assumption establishes a relation between the (properties of the) observations in $\bbX$ and the underlying graph $\ccalG$.
In particular, we consider that the columns of $\bbX$ are (independent) realizations of a GMRF with zero mean and GSO $\bbS$.
While other models relating the graph signals with the unknown supporting graph exist, we focus on GMRF due to its flexibility, solid statistical foundations, and wide adoption within the network science community. 
Nonetheless, in \cref{S:BeyondGMRF} we discuss how to generalize our approach to models beyond GMRF.
\end{itemize}

The goal of this section is to formulate the motif-based graph learning problem rigorously (Problem 1) and postulate an associated constrained optimization problem that leverages the information in $\bbX$ and the previous assumptions to generate as solution the desired $\bbS$.
To that end, we need to describe in more detail our approach to assess motif similarity (remainder of this section and Section \ref{S:density_to_spectrum}) and then set a formulation combining motif similarity with the GMRF topology estimation framework (Section \ref{S:graph_learning_optimization_motifs}). The first step is to describe the structural properties of a graph $\ccalG$ in terms of the density of rooted balls, or motifs.
A rooted graph is simply a graph with a special labeled node, denoted by a tuple $(\ccalG,\rho)$.
If $(\ccalG,\rho)$ is such that each node in $\ccalG$ is in the $r$-hop neighborhood of the root $\rho$, we say that it is a rooted $r$-ball.
For a given integer radius $r\geq 0$, a graph $\ccalG=(\ccalV,\ccalE)$ yields a family of rooted $r$-balls.
For each node $i\in\ccalV$, consider the induced subgraph of the $r$-hop neighborhood of $i$.
Then, treating $i$ as the root, this yields a rooted $r$-ball ``centered'' at $i$, which we denote as $V_r(\ccalG,i)$.

\begin{figure}
    \centering
    \resizebox{.95\linewidth}{!}{\begin{tikzpicture}
  \tikzset{
    bigcirc/.style={
      circle,
      inner sep=0pt,
      text width=5mm,
      align=center,
      draw=black,
    }
  }

  \tikzset{
    smallcirc/.style={
      circle,
      inner sep=0pt,
      text width=3mm,
      align=center,
      draw=black,
    }
  }

  
  \foreach \l/\x/\y/\color in {A_1/6.5/-0.5/blue, A_2/7.5/-1.0/white,
    B_1/9.5/-0.5/red, B_2/9/-1.0/white, B_3/10/-1.0/white} {
    \node (\l) at (\x,\y) [smallcirc,fill=\color,draw] {};
  }

  \foreach \i/\j in {A_1/A_2,
    B_1/B_2, B_1/B_3} {
    \path (\i) edge (\j);
  }
  
  
  \foreach \l/\x/\y/\color in {G_1/12/0/red, G_2/11.5/-0.5/white, 
    G_3/12.5/-0.5/red, G_4/11/-1/blue, G_5/11.5/-1/blue, 
    G_6/12/-1/blue, G_7/12.5/-1/blue, G_8/13/-1/blue} {
    \node (\l) at (\x,\y) [smallcirc,fill=\color,draw] {};  
  }
  
  \foreach \i/\j in {G_1/G_2, G_1/G_3,
    G_2/G_4, G_2/G_5, G_2/G_6,
    G_3/G_7, G_3/G_8} {
    \path (\i) edge (\j);  
  }
  

  \foreach \x/\idx in {7/1, 9.5/2} {
    \node[below] at (\x,-1.25) {$\alpha^{(\idx)}_{1}$};
  }
  
  \node[below] at (12,-1.25) {$\ccalG$};

\end{tikzpicture}}
    \caption{Two rooted $1$-balls $\alpha^{(1)}_1,\alpha^{(2)}_1$ and a graph $\ccalG$.
    Each node of $\ccalG$ is colored to indicate if the rooted $1$-ball centered at that node is isomorphic to $\alpha^{(1)}_1$ (blue), $\alpha^{(2)}_1$ (red), or neither (white).
    From this, we can see that $\tau_1(\alpha^{(1)}_1,\ccalG)=\frac{5}{8}$ and $\tau_1(\alpha^{(2)}_1,\ccalG)=\frac{1}{4}$.}
    \label{fig:motifexample}
\end{figure}
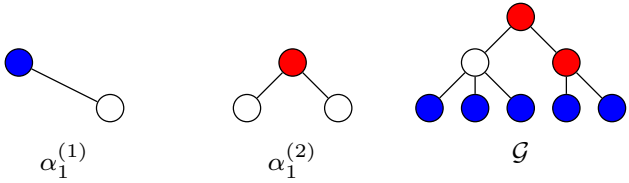

Then, for a given motif $\alpha_r$, we define the \emph{rooted motif density} as
\begin{equation}\label{E:subgraph_dens}
    \tau_r(\alpha_r,\ccalG)=\frac{1}{N}\left|\{i=1,...,N:V_r(\ccalG,i)\cong\alpha_r\}\right|,
\end{equation}
where $V_r(\ccalG,i)\cong\alpha_r$ denotes isomorphism of rooted $r$-balls, i.e., graph isomorphism with the extra condition that the roots coincide.
Simply put, the quantity $\tau_r(\alpha_r,\ccalG)$ measures the frequency with which a specific motif $\alpha_r$ appears in the graph $\ccalG$ by computing the proportion of rooted $r$-balls in $\ccalG$ that are isomorphic to $\alpha_r$.
We illustrate this in \cref{fig:motifexample}.

Based on \eqref{E:subgraph_dens}, we can compare the similarity between two graphs in terms of their motif densities. With this notation at hand, we formalize the graph learning problem introduced at the beginning of the section.
\begin{problem}\label{P:nti_motifs}
Let $\ccalG$ be an unknown graph with node set $\ccalV$, $N=|\ccalV|$ and GSO $\bbS\in\reals^{N \times N}$. Furthermore, i)  let  $\tilde{\ccalG}$ be a reference graph with node set $\tilde{\ccalV}$, $\tilde{N}=|\tilde{\ccalV}|$ and GSO $\tbS\in\reals^{\tilde{N} \times \tilde{N}}$ and ii) let $\bbX\in\reals^{N \times M}$ be a set of $M$ graph signals defined over $\ccalG$. Our goal is to use $\bbX$ and $\tilde{\ccalG}$ to find the underlying graph structure encoded in $\bbS$ under the assumptions that:
    
    \vspace{1mm}
    \noindent\textbf{(AS1a)} Graphs $\ccalG$ and $\tilde{\ccalG}$ have nodes with degree at most $D$.
    
    \vspace{1mm}
    \noindent\textbf{(AS1b)} Graphs $\ccalG$ and $\tilde{\ccalG}$ present a similar density of motifs, so that $|\tau_r(\alpha_r^{(k)},\ccalG)-\tau_r(\alpha_r^{(k)},\tilde{\ccalG})|\leq\epsilon$ for every $k$, with $\epsilon$ being a small positive number, $r\in(0,R]$, and $\{\alpha_r^{(k)}\}_{k=1}^K$ being the set of all (isomorphisms of) rooted $r$-balls inside the graph $\ccalG$. 
    
        \vspace{1mm}
    \noindent\item \textbf{(AS2)} The columns of $\bbX$ are $M$ independent realizations of a GMRF with zero mean and GSO $\bbS$ [cf. \eqref{E:gaussian_pdf_S}].
\end{problem}
%

Examining the proposed motif-related assumptions, we note that \textbf{(AS1a)} ensures that there are finitely many possible $r$-balls on a given graph, which will be used in the derivation presented in the following section.
On the other hand, \textbf{(AS1b)} provides prior information about the density of motifs of the sought graph based on a structurally similar reference graph.
From the definition of rooted motif density in \eqref{E:subgraph_dens}, we can observe that
$\tau_r(\alpha_r,\ccalG)$ is an expectation of the frequency with which the motif $\alpha_r$ appears in the graph $\ccalG$.
Moreover, since this expectation is computed locally at each node, \textbf{(AS1b)} endows the inference problem with some interesting properties.
First, it allows us to compare graphs of different sizes, something that was non-trivial in other works where the graph similarity promoted graphs with similar supports~\cite{danaher2014joint,rey2021joint}.
Also, note that assuming that two graphs have similar densities of motifs is a laxer requirement than assuming they have similar supports.
Second, we do not require to know the whole graph $\tilde{\ccalG}$ since we can approximate its associated motif density through a smaller subgraph, so knowing a sampled version of $\tilde{\ccalG}$ suffices.

Unfortunately, despite all the attractive properties previously discussed, the rooted motif density is intrinsically a combinatorial metric and, as a result, its direct incorporation into an optimization framework leads to an NP-hard problem. The next section lays down an approach to circumvent this issue.

\subsection{From similar densities of motifs to spectral distributions}\label{S:density_to_spectrum}
Our aim is to find an alternative approach to take advantage of the graph similarity specified in \textbf{(AS1b)} without falling into an NP-hard combinatorial problem.
To that end, we start by noting that, due to the nature of the GSO, the diagonal entries of $\bbS^r$ are strictly dictated by the $r$-balls centered at each node.
Furthermore, since $\tr(\bbS^r)=\tr(\bbLambda^r)$, it seems evident that the density of motifs is closely related to the eigenvalues of the GSO, collected in the $N \times N$ diagonal matrix $\bbLambda=\diag(\bblambda)$.
This suggests that the spectra of two graphs with similar densities of motifs should be similar.

Motivated by the previous discussion, we encode the similar density of motifs between two graphs by means of test functions applied to the spectral distribution of the graphs.
Let $\bblambda\in\reals^N$ denote the vector containing the eigenvalues of $\bbS$, and denote its associated empirical spectral density function as $\mu_{\bblambda}$, with $\mu_{\bblambda}(\lambda_i)$ quantifying the multiplicity of the $i$th eigenvalue normalized by the number of nodes in $\bbS$.
Indeed, $\mu_{\bblambda}$ is (formally) a probability distribution on $\reals$.
Then, for any continuous function $g:\reals\to\reals$ compute the Lebesgue integral
\begin{equation}\label{E:test_function}
    c_g(\bblambda) = \int g(\lambda) \;d\mu_{\bblambda}(\lambda) = \frac{1}{N}\sum_{i=1}^N g(\lambda_i),  
\end{equation}
where the last equality follows from $\bbS$ having a discrete spectrum.
With these definitions in place, the following result shows that if $\bbS$ and $\tbS$ have similar densities of motifs, then $c_g(\bblambda)$ and $c_g(\tblambda)$ are close.

\begin{theorem}\label{T:spectral_closeness}
    Let $\bblambda\in\reals^N$ and $\tblambda\in\reals^{\tilde{N}}$ denote the eigenvalues of the GSOs of the graphs $\ccalG$ and $\tilde{\ccalG}$.
    For any continuous test function $g$, under \textbf{(AS1a)} and \textbf{(AS1b)}, it follows that
    \begin{equation}\label{E:sim_test_functs}
        |c_g(\bblambda) - c_g(\tblambda)| \leq \delta,
    \end{equation}
    where $c_g(\cdot)$ is given in \eqref{E:test_function}, and $\delta\geq0$ is a constant dependent only on $g$, $r$, $D$, and $\epsilon$ [cf. \textbf{(AS1a), (AS1b)}], such that $\delta\to 0$ as $\epsilon\to 0$.
\end{theorem}

The proof of the theorem is provided in Appendix~\ref{A:spectral_closeness}.
In a nutshell, the proof shows that for any continuous test function $g$, the quantity $c_g(\bblambda)$ can be expressed as the expected value of some continuous function depending only on rooted $r$-balls.
Hence, since \textbf{(AS1b)} implies that $\ccalG$ and $\tilde{\ccalG}$ present similar densities of motifs (rooted $r$-balls), by the continuity of this function involving $r$-balls, we show that $c_g(\bblambda)$ and $c_g(\tblambda)$ are close.

\Cref{T:spectral_closeness} allows us to reduce the similarity of motif densities between two graphs to a comparison of an appropriate test function $g$ applied to their empirical spectral densities.
Although the quantity $c_g(\bblambda)$ is less expressive than the motif densities $\tau_r(\alpha_r^{(k)},\ccalG)$ in describing the structural properties of a graph, it bypasses the combinatorial issues of computing the precise motif densities.
As we will see in the following section, this tradeoff is worthwhile, as it allows for seamless incorporation into network topology inference methods.

\subsection{Graph motif-enhanced optimization for GMRF learning}\label{S:graph_learning_optimization_motifs}
Suppose for now that we ignore the assumptions \textbf{(AS1a)} and \textbf{(AS1b)}. Leveraging \textbf{(AS2)} and the PDF in \eqref{E:gaussian_pdf_S}, we have that the likelihood of the joint observation of the $M$ signals in $\bbX=[\bbx_1,...,\bbx_M]$ is 
$\prod_{m=1}^M (2\pi)^{-N/2} \det(\bbS)^{1/2} \exp\left(-\frac 1 2 \bbx_m^\mathrm{T}{\bbS}\bbx_m\right)$. Upon adopting a maximum likelihood (ML) approach, exploiting the monotonicity of the log function, and using the observations in $\bbX$ to build the empirical covariance matrix $\hbC=\frac{1}{M}\sum_{m=1}^M\bbx_m\bbx_m^\top$, the matrix $\bbS$ can be estimated as   
\begin{alignat}{2}\label{E:graphical_lasso} 
    \!\!&\! \min_{\bbS} \
    && \tr(\hbC\bbS) \!-\! \log\det(\bbS)   \nonumber \\ 
    \!\!&\! \mathrm{s.t}: && 
    \bbS \succeq 0,
\end{alignat} 
with the constraint $\bbS\succeq 0$ guaranteeing that the precision matrix is positive semidefinite and that the $\log\det$ function in the objective is well defined. In the context of GMRF, a widely adopted approach is to augment the objective in \eqref{E:graphical_lasso} with a sparsity promoting regularizer $\lambda \|\bbS\|_1$, giving rise to the celebrated graphical lasso algorithm \cite{lauritzen1996graphical,friedman2008sparse,ravikumar2011high}. In the previous, $\lambda>0$ controls the level of sparsity and $\|\bbS\|_1$ denotes the $\ell_1$ norm of the vectorization of the matrix $\bbS$. On top of augmenting the ML formulation with an $\ell_1$ norm, other graph learning approaches incorporate topological conditions by considering a set of feasible GSOs $\ccalS$ and augmenting the formulation in \eqref{E:graphical_lasso}  with the constraint $\bbS\in\ccalS$~\cite{lake2010discovering,zhao2019optimization}.

Hence, the key to our approach is to formulate a modified version of the ML estimation in \eqref{E:graphical_lasso} capable of exploiting the availability of the reference graph $\tilde{\ccalG}$ and the results in  \cref{T:spectral_closeness}.
More specifically, we encode the fact of $\ccalG$ and $\tilde{\ccalG}$ having similar densities of motifs by leveraging \eqref{E:sim_test_functs} and, as a result, approach Problem~\ref{P:nti_motifs} through the following non-convex optimization program:
\begin{alignat}{2}\label{E:noncvx_problem} 
    \!\!&\! \min_{\bbS,\bbV,\bblambda} \
    && \tr(\hbC\bbS) \!-\! \log\det(\diag(\bblambda)) + \alpha\|\bbS\|_1 \nonumber \\
    \!\!&\!  && + \frac{\beta}{2}\|\bbS - \bbV\diag(\bblambda)\bbV^\top\|_F^2  \nonumber \\ 
    \!\!&\! \mathrm{s.t}: && 
    |c_g(\bblambda) - c_g(\tblambda)| \leq \delta, \;\;  \bbS \in \ccalS, \;\; \bbV^\top\bbV=\bbI.
\end{alignat} 
Note that this alternative formulation for learning GMRFs is amenable to constraints involving the spectrum of $\bbS$.
Also, recall that $\tblambda$ denotes the eigenvalues related to the reference graph, so $c_g(\tblambda)$ is a known constant.

We refer to the first constraint in \eqref{E:noncvx_problem} as the similarity constraint because, following \cref{T:spectral_closeness}, it stems from the fact that $\ccalG$ and $\tilde{\ccalG}$ have similar motif densities.
At a high level, it promotes desired properties over the eigenvalues of $\bbS$ by ensuring that evaluating the empirical spectral distribution of $\bbS$ and $\tbS$ with a common test function $g$ yields a similar value.
If we are interested in further reducing the size of the feasible set, it is possible to simultaneously employ several test functions $\{g_j\}_{j=1}^J$ resulting in the associated set of functions $\{c_{g_j}\}_{j=1}^J$.
We can trivially modify the program in \eqref{E:noncvx_problem} to include a similarity constraint for each function $c_{g_j}$. 
When several constraints are included, we face a trade-off between the improvement in the estimation of $\bbS$ and the additional complexity of enlarging the set of constraints.
In the remainder of the paper, we assume that a single similarity constraint is used, and leave the (optimal) combination of multiple constraints as a future research direction.

The optimization framework introduced in \eqref{E:noncvx_problem} estimates separately the GSO $\bbS$ from its eigendecomposition $\bbV\diag(\bblambda)\bbV^\top$, including a Frobenius-norm penalty in the objective function to encourage that $\bbS$ and $\bbV\diag(\bblambda)\bbV^\top$ stay close.
Dealing with $\bbV$ and $\bblambda$ as explicitly separated optimization variables allows us to incorporate constraints involving the spectrum of the graph. While this sacrifices convexity, the selected approach is amenable to designing an efficient iterative algorithm, as detailed in Section~\ref{S:algorithm}.
Consideration of graph eigenvalues as explicit optimization variables in the context of graph learning has been explored in, e.g., \cite{segarra2017network} and \cite{kumar2019structured}.
In \cite{segarra2017network} the eigenvectors were considered to be given. Meanwhile, in \cite{kumar2019structured} they consider that $\bbS=\bbL$ and the (convex) spectral constraints are mainly concerned with relatively simple conditions, such as bounding the minimum and maximum value of non-zero elements in $\bblambda$ or selecting the number of connected components (number of zero eigenvalues).
Differently, the similarity constraints considered in this paper are more involved, lead to non-convex formulations and emanate from the assumption that two graphs present similar densities of motifs. These differences will be further investigated in the numerical experiments presented in Section~\ref{S:experiments}.

Capturing more complex prior information about (the spectrum of) $\bbS$ comes at the cost of employing non-convex constraints. However, since the optimization in \eqref{E:noncvx_problem} was already non-convex, it does not fundamentally change the complexity of the problem. This is further discussed in the following section where a convex-approximation approach to handle the similarity constraints is introduced.

\section{Convex relaxation for the similarity constraints}\label{S:convex_relaxation}
Solving the optimization problem introduced in \eqref{E:noncvx_problem} is a challenge due to its non-convexity, stemming from the bilinear terms involving $\bbV$ and $\bblambda$, the orthogonality of $\bbV$, and the similarity constraint.
The bilinear terms and the orthogonality constraint can be dealt with by implementing an alternating optimization scheme and leveraging results from optimization over manifolds~\cite{absil2009optimization}, respectively. However, dealing with the \emph{similarity constraint} requires further elaboration.

To analyze the curvature of the similarity constraint, we start by noting that $|c_g(\bblambda)-c_g(\tblambda)| \leq \delta$ is a composition of functions, an operation that is non-convex in general~\cite{boyd2004convex}.
We also observe that the convexity of $c_g$ is determined by the convexity of the test function $g$.
Then, due to the presence of the absolute value, the similarity constraint will only be convex when the considered test function $g$ is affine.

According to the definition of the function $c_g(\cdot)$ provided in \eqref{E:test_function}, it follows that any affine function $g(x)=ax+b$ with $a,b\in\reals$ delimits the same feasible set independently of the values of $a$ and $b$.
Thus, we select the affine function $g(x)=x$, which results in the similarity constraint
\begin{equation}\label{E:tr_constraint}
    \left|\frac{1}{N}\sum_{i=1}^N \lambda_i - C\right| = \left|\frac{1}{N}\tr(\bbS) - C\right|\leq \delta,
\end{equation}
where the constant $C:=c_g(\tblambda)$ encodes the value of the test function evaluated over the known reference graph.
A closer inspection reveals that, when $C=1$ and $\delta=0$, \eqref{E:tr_constraint} is equivalent to $\tr(\bbS)=N$, a common constraint used to fix the scale of the GSO when learning the graph topology~\cite{dong2016learning}.
That is to say, the constraint $\tr(\bbS)=N$ represents a particular case of the similarity constraints put forth in this paper.
Moreover, using \eqref{E:tr_constraint} as a constraint incorporates information about the true scale of the graph, avoiding the scale ambiguity inherent to most network topology inference approaches.
Indeed, we observe in \cref{S:experiments} that this general approach reduces the scale ambiguity of the estimated GSO. 

Nonetheless, using a linear test function might not be enough to capture more complex relations between the spectral distributions of $\bbS$ and $\tbS$.
We tackle this issue below by discussing a convex alternative to leverage more general classes of test functions.

\subsection{Convex relaxation for convex or concave test functions}
Since our goal is to develop a convex relaxation for the similarity constraint defined in \eqref{E:sim_test_functs}, we can focus on either convex or concave test functions $g$ without loss of generality.
Therefore, we start our discussion by proposing a convex relaxation under the assumption that $g$ is \emph{concave}.

We already discussed that the similarity constraint $|c_g(\bblambda)-c_g(\tblambda)| \leq \delta$ is non-convex due to the composition of the absolute value and the function $c_g(\bblambda)-c_g(\tblambda)$.
Then, the first step towards obtaining a convex surrogate consists of decomposing the similarity constraint into
\begin{align}\label{E:split_similarity}
    &c_g(\bblambda) \leq c_g(\tblambda) + \delta
    &c_g(\bblambda) \geq c_g(\tblambda) -\delta,
\end{align}
where the left and the right constraints are respectively concave and convex due to the concavity of $g$.

%
The pair of constraints in \eqref{E:split_similarity} determines a feasible set equivalent to the one determined by our original similarity constraint based on the composition of functions.
Hence, we replace the optimization problem in \eqref{E:noncvx_problem} with its equivalent form
\begin{alignat}{2}\label{E:split_constraint}
    \!\!&\! \min_{\bbS,\bbV,\bblambda} \
    && \tr(\hbC\bbS) \!-\! \log\det(\diag(\bblambda)) + \alpha\|\bbS\|_1 \; \nonumber \\
    \!\!&\!  && + \frac{\beta}{2}\|\bbS - \bbV\diag(\bblambda)\bbV^\top\|_F^2 + \gamma c_g(\bblambda) \nonumber \\ 
    \!\!&\! \mathrm{s.t}: && 
    c_g(\bblambda) \geq c_g(\tblambda)-\delta, \;\;  \bbS \in \ccalS, \;\; \bbV^\top\bbV=\bbI.
\end{alignat}
Here, we kept the convex term from \eqref{E:split_similarity} as a constraint while the concave term is used to augment the objective function.
Note that, from the perspective of duality theory, any constraint can be equivalently expressed as a regularization term in the objective function with a non-negative parameter (here denoted as $\gamma$) playing the role of the dual variable.

Even though the objective function of \eqref{E:split_constraint} is still non-convex due to the presence of convex and concave terms, now the optimization problem can be efficiently solved by an MM approach~\cite{sun2016majorization}.
Based on the MM framework, we consider an iterative linear upper bound to the function $c_g(\bblambda)$ leading to a \emph{convex iterative} algorithm that approximates the similarity constraint.
Because $c_g(\bblambda)$ is concave, a suitable upper bound is provided by
\begin{equation}
    u(\bblambda,\bblambda^{(t-1)}) = \nabla c_g(\bblambda^{(t-1)})^\top\bblambda,
\end{equation}
which is the first-order approximation of the Taylor series of $c_g$ centered at the solution of the previous iteration $\bblambda^{(t-1)}$.
Note that we have omitted the terms that do not involve the variable $\bblambda$ since they are constants in the optimization problem.

Intuitively, the original non-convex similarity constraint  $|c_g(\bblambda)\!-\!c_g(\tblambda)| \!\leq\! \delta$ ensured that
\begin{equation}\label{E:true_interval}
    c_g(\bblambda)\in [c_g(\tblambda)-\delta,\; c_g(\tblambda)+\delta]
\end{equation}
for any feasible $\bblambda$.
Now, with the proposed convex relaxation based on the MM algorithm, the feasible set is modified as follows.
First, the convex constraint $c_g(\bblambda) - c_g(\tblambda) \geq \delta$ in \eqref{E:split_constraint} ensures that
\begin{equation}
    c_g(\bblambda)\in [c_g(\tblambda)-\delta, \infty].
\end{equation}
Then, successively minimizing the upper bound $u(\bblambda,\bblambda^{(t-1)})$ brings the value of $c_g(\bblambda)$ closer to $c_g(\tblambda)-\delta$, the minimum value inside the feasible set.
Thus, the value of $\gamma$ is chosen to promote that $c_g(\tblambda)$ is inside the interval defined in \eqref{E:true_interval}.
This process can be interpreted as starting with a loose constraint for the maximum value of $c_g(\tblambda)$ that gets tightened as the iterative algorithm converges.
All the details about the specific implementation of the convex iterative algorithm that solves \eqref{E:split_constraint} are provided in Section~\ref{S:algorithm}.

The last step is to discuss the formulation for \emph{convex} functions $g$, which lead to a convex $c_g(\bblambda)$. Using an approach analogous to that for the concave case, from the two constraints in \eqref{E:split_similarity} we incorporate the convex one into the graph-related optimization. This entails replacing $c_g(\bblambda)\geq c_g(\tblambda)-\delta$ with $c_g(\bblambda)\leq c_g(\tblambda)+\delta$ in \eqref{E:split_constraint}. Additionally, since for the convex case we are interested in maximizing $c_g(\bblambda)$, we replace $\gamma c_g(\bblambda)$ with $-\gamma c_g(\bblambda)$ in the objective of \eqref{E:split_constraint} and employ an MM approach to minimize a linear upper bound of $-\gamma g(\bblambda)$. 

To summarize, following an MM approach we obtain a convex relaxation for the similarity constraint for every test function $g$ that is differentiable and either convex or concave.
Next, we present the specific iterative algorithm that simultaneously deals with the MM relaxation, the bilinear terms, and the orthogonality constraints.

\section{Algorithmic implementation}\label{S:algorithm}
We solve the network topology inference task presented in Problem~\ref{P:nti_motifs} by developing an iterative algorithm that solves~\eqref{E:split_constraint}.
To that end, we combine an alternating optimization approach that decouples the bilinear terms involving $\bblambda$ and $\bbV$ via MM while incorporating the convex relaxation of the similarity constraint.
The resulting algorithm falls into the family of Block Successive Upper bound Minimization (BSUM)~\cite{hong2015unified}.
This class of algorithms blend techniques from MM and alternating optimization, and they converge to a stationary point under mild conditions.

Our proposed BSUM algorithm solves \eqref{E:split_constraint} by updating the optimization variables $\bbS$, $\bbV$, and $\bblambda$ in \emph{three} separated \emph{steps}.
At each step we optimize over one of the optimization variables while the rest remain fixed, procuring simpler problems that can be solved efficiently.
Then, for a maximum number of $T$ iterations, the following steps are computed at each iteration $t=0,1,...,T$.

\vspace{2mm}\noindent
\textbf{Step 1.} The first step estimates the block of variables represented by $\bbS$ while the rest remain fixed.
This results in the convex optimization problem given by 
\begin{alignat}{3}\label{E:step1} 
    \!\!&\!\ \bbS^{(t+1)} = && \argmin_{\bbS} \
    && \tr(\hbC\bbS) \!+\! \alpha\|\bbS\|_1  \!+\! \frac{\beta}{2}\|\bbS \!-\! \bbV^{(t)}\bbLambda^{(t)}\bbV^{(t)^\top}\|_F^2  \nonumber \\
    \!\!&\!  && \mathrm{s.t}: && \!\!\! \bbS \in \ccalS,
\end{alignat} 
where $\bbLambda^{(t)}=\diag(\bblambda^{(t)})$. The optimization in \eqref{E:step1} is a combination of linear and (convex) quadratic terms, that can be handled by a number of algorithms. The one we advocate here is a straightforward adaptation of the approach presented in~\cite{kumar2019structured}.
To that end, let $\bbH$ be a matrix of signed ones matching the sign of the entries of $\bbS$ such that $\|\bbS\|_1=\tr(\bbS\bbH)$, and hence, $\tr(\hbC\bbS)+\alpha\|\bbS\|_1=\tr(\bbK\bbS)$, where $\bbK=\hbC+\bbH$.
Also, define the linear operator $\bbcalS:\bbs\in\reals_+^{N(N-1)/2}\to\bbcalS\bbs\in\reals^{N \times N}$ that maps the vector $\bbs$ into the matrix $\bbS=\bbcalS\bbs$ satisfying the constraints in $\ccalS$, and denote the adjoint linear operator of $\bbcalS$ as $\bbcalS^*:\bbY\in\reals^{N\times N}\to\bbcalS^*\bbY\in\reals^{N(N-1)/2}$.
Then, we efficiently approximate \eqref{E:step1} by solving 
\begin{equation}\label{E:step1_simplified}
    \bbs^{(t+1)}=\left(\bbs^{(t)}-\frac{1}{\|\bbcalS\|_2^2}(\bbcalS^*(\bbcalS\bbs^{(t)})-\bbz) \right)^+,
\end{equation}
where $\bbz \!=\! \bbcalS^*\!(\bbV^{(t)}\!\bbLambda^{(t)})\!\bbV^{(t)^\top} \!-\! \beta^{-1}\bbK)$, $(a)^+=\max(a,0)$, and $\|\bbcalS\|_2$ denotes the operator norm.
Finally, we update $\bbS^{(t+1)}$ as $\bbS^{(t+1)}=\bbcalS\bbs^{(t+1)}$.

The derivation of the solution presented in \eqref{E:step1_simplified} from the initial problem \eqref{E:step1} is provided in Appendix~\ref{A:step1} for completeness.

\vspace{2mm}\noindent
\textbf{Step 2.}
The second step estimates the block of variables $\bbV$ while the others remain fixed.
Ignoring the constant terms, the resulting optimization problem is given by
\begin{alignat}{3}\label{E:step2_a}
    \!\!&\! \bbV^{(t+1)} = && \;\; \argmin_{\bbV} \;\; && \frac{\beta}{2}\|\bbS^{(t+1)}-\bbV\bbLambda^{(t)}\bbV^\top\|_F^2 \nonumber \\
    \!\!&\! && \;\; \mathrm{s.t}: \;\; && \bbV^\top\bbV = \bbI,
\end{alignat}
which can be equivalently rewritten as
\begin{alignat}{3}\label{E:step2_b} 
    \!\!&\! \bbV^{(t+1)} = && \;\; \argmax_{\bbV} \;\; && \tr(\bbV^\top\bbS^{(t+1)}\bbV\bbLambda^{(t)}) \nonumber \\
    \!\!&\!  && \;\; \mathrm{s.t}: \;\; && \bbV^\top\bbV = \bbI.
\end{alignat}

We note that the orthogonality constraint implies that the optimization variables $\bbV$ belong to the Stiefel manifold.
This is a well-known optimization problem and, as explained in \cite[Chapter 4.8]{absil2009optimization}, it follows that the solution to \eqref{E:step2_b} is setting $\bbV^{(t+1)}$ to the eigenvectors of $\bbS^{(t+1)}$.

\vspace{2mm}\noindent
\textbf{Step 3.}
The last step estimates the block of variables $\bblambda$ while the others remain fixed.
The resulting optimization problem after ignoring the constant terms can be compactly written as
\begin{alignat}{3}\label{E:step3_simplified} 
    \!\!&\!\ \bblambda^{(t+1)}= && \;\argmin_{\bblambda}
    && -\sum_{j=1}^N \log(\lambda_j) +  \frac{\beta}{2}\|\bblambda-\hblambda\|_2^2 + \gamma u(\bblambda, \bblambda^{(t)}) \; \nonumber \\ 
    \!\!&\! && \; \mathrm{s.t}: && \!\!\! c_g(\bblambda) \geq c_g(\tblambda) - \delta,
\end{alignat}
where $u(\bblambda,\bblambda^{(t)})$ denotes the linear majorization of $c_g(\bblambda)$ at $\bblambda^{(t)}$, and the vector $\hblambda$ collects the elements on the diagonal of $\bbV^{(t+1)^\top}\bbS^{(t+1)}\bbV^{(t+1)}$, which are the eigenvalues of $\bbS^{(t+1)}$.
Recall that combining the inequality constraint and the minimization of the upper bound $u(\bblambda,\bblambda^{(t)})$ incorporates the prior information about the distribution of the graph spectrum.
Moreover, \eqref{E:step3_simplified} assumes that the test function $g$ is concave, but, as explained in Section \ref{S:convex_relaxation}, the formulation can be easily modified to account for a convex $g$.

\begin{algorithm}[tb]
\SetKwInput{Input}{Input}
\SetKwInOut{Output}{Output}
\Input{$\hbC$, $c_g(\tblambda)$}
\Output{$\hbS$.}
\SetAlgoLined
Initialize $\bbS^{(0)}$, $\bbs^{(0)}$, $\bblambda^{(0)}$, and $\bbV^{(0)}$. \\
\For{$t=1$ \KwTo $T$}{
    Set $\bbs^{(t+1)}$ as in \eqref{E:step1_simplified}. \\
    $\bbS^{(t+1)}=\bbcalS\bbs^{(t+1)}$. \\
    Set $\bbV^{(t+1)}$ as the eigenvectors of $\bbS^{(t+1)}$. \\
    Set $\bblambda^{(t+1)}$ as the solution to \eqref{E:step3_simplified}. \\
}
$\hbS=\bbS^{(T)}$
\caption{Graph learning from similarity constraints.}
\label{A:NTI_tight}
\end{algorithm}

The overall procedure is summarized in Algorithm~\ref{A:NTI_tight}.
Analyzing its computational complexity, we observe that Step~1 requires a moderate number of operations while the complexity of Step~2 is $\ccalO(N^3)$ because it computes the eigendecomposition of $\bbS$.
Regarding Step~3, directly solving the optimization problem in \eqref{E:step3_simplified} would result in a computational complexity of $\ccalO{(N^{3.5})}$.
However, because the problem is strictly convex and separable for each optimization variable $\lambda_j$, it can be solved efficiently resulting in a much smaller computational complexity.
As a result, the most expensive operation in practice is the eigendecomposition performed in the second step, and hence, the complexity of the overall algorithm is $\ccalO(N^3)$.
We emphasize that this is fairly efficient since the optimization problem involves $\ccalO(N^2)$ variables. 
Furthermore, recall that the assumptions in Problem~\ref{P:nti_motifs} involve relating the motif densities of two different graphs, which is a challenging NP-hard combinatorial problem.


Another key aspect of the proposed BSUM algorithm is its convergence to a stationary point, which is formally stated in the following proposition.
\begin{proposition}
    Let $\ccalY^*$ denote the set of stationary points of \eqref{E:split_constraint}.
    Then, the sequence $(\bbS^{(t)}, \bbV^{(t)}, \bblambda^{(t)})$ generated by Algorithm~\ref{A:NTI_tight} converges to a stationary point in $\ccalY^*$ as $t\to\infty$.
\end{proposition}

To prove the convergence of our algorithm, we leverage the results in \cite{hong2015unified} and \cite{fu2017scalable}. To be more specific, conditions under which BSUM algorithms converge to a stationary point were identified in \cite[Th.~1b]{hong2015unified}. However, the original result in \cite{hong2015unified} did not consider formulations with non-convex constraints, and this is relevant in our setup because the optimization problem in Step~2 includes the non-convex orthogonality constraint $\bbV^\top\bbV = \bbI$. Fortunately, in the context of tensor decompositions, \cite{fu2017scalable} proved that the sequence generated by BSUM algorithms still converges when considering orthogonality constraints like the one in Step 2. 
As a result, leveraging \cite{fu2017scalable}, we can prove the claim in Proposition 1 by showing that the our problem satisfies the original conditions identified in \cite[Th.~1b]{hong2015unified}. To be precise, upon denoting the objective function in \eqref{E:split_constraint} as $\phi(\bbS,\bbV,\bblambda)$, we have that: (i) the objective functions in \eqref{E:step1}, \eqref{E:step2_b}, and \eqref{E:step3_simplified} are upper bounds of $\phi(\bbS,\bbV,\bblambda)$\blue{\footnote{To be rigorous, when stating that the objective functions of the steps 1, 2 and 3 are upper bounds of $\phi(\bbS,\bbV,\bblambda)$ we are also considering the constant terms omitted in the optimization problems \eqref{E:step1}, \eqref{E:step2_b}, and \eqref{E:step3_simplified}.}} ; (ii) the level set $\{(\bbS,\bbV,\bblambda)~|\phi(\bbS,\bbV,\bblambda) \leq \phi(\bbS^{(0)},\bbV^{(0)},\bblambda^{(0)})\}$ is compact; (iii) the optimization problems in Step 1 and Step 3 are strictly convex; and (iv) the non-smooth components of $\phi(\bbS,\bbV,\bblambda)$ only involve the variables in $\bbS$.
As a result, the conditions specified in \cite[Th.~1b]{hong2015unified} are met and, invoking \cite[Th.~1b]{hong2015unified} and \cite{fu2017scalable}, it follows that the solution of our algorithm converges to a stationary point.

\section{Beyond GMRFs}\label{S:BeyondGMRF}
To simplify exposition and promote clarity, our discussion has been focused on addressing the motif-similarity graph-learning design for the conditions outlined in \cref{P:nti_motifs}.
However, as pointed out at different points of the manuscript, our approach can be used under more general circumstances than those considered this far.
Three generalizations particularly appealing are: (i) having access to more than one reference graph $\tilde{\ccalG}_r$; (ii) having access to the actual spectral density function as $\mu_{\bblambda}$ in lieu of $\tilde{\ccalG}$; and (iii) considering more general models than a GMRF to represent the relation between the signals $\bbX$ and the GSO $\bbS$. Next, we briefly discuss the modifications to the optimization in \eqref{E:split_constraint} required to account for these generalizations.  

Starting with the first generalization, let us suppose that we have access to $R$ reference graphs, denoted as $\{\tilde{\ccalG}_r\}_{r=1}^R$. Assuming that the sought graph $\ccalG$ is similar to the graphs in $\{\tilde{\ccalG}_r\}_{r=1}^R$ requires only considering the set of constraints [cf.~\eqref{E:split_similarity}] 
\begin{align}\label{E:split_similarity_multiple_ref_graphs}
    &c_g(\bblambda) \leq c_g(\tblambda_r) + \delta_r
    &c_g(\bblambda) \geq c_g(\tblambda_r) -\delta_r, \;\; 
\end{align}
for all $r$.
Here, $\tblambda_r$ denotes the eigenvalues of the $r$-th reference graph and the value of $\delta_r$ can be selected based on prior information on the similarity between $\ccalG$ and $\tilde{\ccalG}_r$. If such information does not exist, then $\delta_r$ is set to $\delta$ for all $r$. Moreover, while all the constraints in \eqref{E:split_similarity_multiple_ref_graphs} can be incorporated into \eqref{E:split_constraint}, a more prudent approach is to identify first the most restrictive ones and then augment the constraints (objective) of \eqref{E:split_constraint} only with those.

We might encounter several reference graphs with similar densities of motifs if, e.g., they are samples drawn from a common random graph model.
This leads us to the second generalization, which consists in having access to the desired (true) spectral density function $\mu_{\bblambda}$ associated with the random graph model at hand.
With an eye on real-world applications, the paper has mostly focused on the case where the prior information on the distribution of motifs comes from a reference graph $\tilde{\ccalG}$ and its empirical spectral density function.
However, there may be cases where the actual spectral density function $\mu_{\bblambda}$ is known or, alternatively, where promoting some desired properties over the spectral density is of interest.
The key to designing graph-learning algorithms that handle the knowledge of $\mu_{\bblambda}$ efficiently is to leverage \eqref{E:test_function}, which relates the evaluation of the test functions over the ensemble and the sample distribution.
More specifically, it suffices with replacing the sample estimate $c_g(\tblambda)$ in constraint \eqref{E:split_similarity} with the ensemble estimate $\int g(\lambda) \;d\mu_{\bblambda}(\lambda)$ computed based on $\mu_{\bblambda}$, with no additional changes being required in the optimization.


The third generalization deals with more encompassing models to represent the relation between the observed signals and the sought graph. A meaningful and tractable alternative is to consider that the signals are Gaussian and graph stationary \cite{perraudin2017stationary,marques2017stationary}.
Basically, a zero-mean random graph signal $\bbx$ is said to be stationary in a GSO $\bbS$ if its covariance matrix $\bbC_x=\EE[\bbx\bbx^T]$ can be written as a polynomial of $\bbS$ \cite{marques2017stationary}.
Clearly, GMRFs are a particular instance of graph stationary models, since we have that $\bbC_x=\bbS^{-1}$.
As a result, graph stationarity has been recently used in a number of graph-learning related problems \cite{segarra2017network,shafipour2020online,rey2021robust}.
For the setup at hand, considering that the signals are both Gaussian and graph stationary implies that the eigenvectors of $\bbS$ and those of the precision matrix $\bbTheta\in\reals^{N \times N}$ are the same and, as a result, that the product $\bbS\bbTheta$ is the same as the product $\bbTheta\bbS$. Then, a tractable way to adapt our formulation in \eqref{E:split_constraint} to deal with stationary GMRF signals is to consider the constraint $\bbS\bbTheta=\bbTheta\bbS$, which results in the following optimization problem
\begin{alignat}{2}\label{E:stationarity}
    \!\!&\! \min_{\bbTheta,\bbS,\bbV,\bblambda} \
    && \tr(\hbC\bbTheta) \!-\! \log\det(\bbTheta) + \alpha\|\bbS\|_1 \; \nonumber \\
    \!\!&\!  && + \frac{\beta}{2}\|\bbS - \bbV\diag(\bblambda)\bbV^\top\|_F^2 + \gamma c_g(\bblambda) \nonumber \\ 
    \!\!&\! \mathrm{s.t}: && 
    c_g(\bblambda) \geq c_g(\tblambda)-\delta, \;\;  \bbS \in \ccalS, \;\; \bbV^\top\bbV=\bbI, \nonumber \\
    \!\!&\!   && \bbTheta\bbS=\bbS\bbTheta.
\end{alignat}
Intuitively, rather than promoting a sparse precision matrix such that $\bbTheta=\bbS$, \eqref{E:stationarity} learns a precision matrix $\bbTheta$ that is a polynomial of the sparse GSO.
This less restrictive assumption results in a more flexible graph-learning algorithm capable of handling a larger range of scenarios.
Even though the resulting optimization problem is non-convex, it is amenable to an iterative approach similar to the one presented in \cref{S:algorithm}, but with an additional step for estimating the new optimization variable $\bbTheta$.

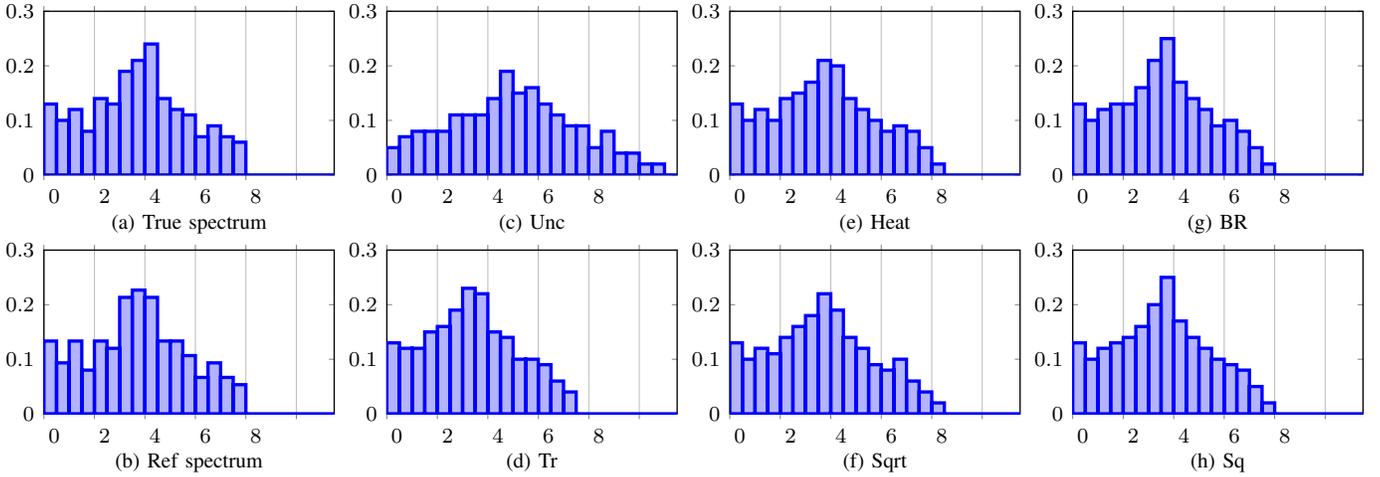
\begin{figure*}[!t]
	\centering
	\pgfplotstableread{data/constraints_hists.csv}\datatable
\pgfplotstableread{data/constraints_hists_ref.csv}\datatableref

\begin{tikzpicture}[baseline]
\begin{groupplot}[
    y post scale=.7,
    group style={group size=4 by 2,
        horizontal sep=0.7cm,
        vertical sep=1cm,},
    width=0.3\textwidth,
    xlabel style={yshift=1mm},
    tickwidth=2pt,
    ybar interval,
    ymin=0, ymax=0.3,
    xmin=0, xmax=11.5,
    xtick={0,2,...,12},
    xticklabel style={anchor=north east},
    ]]
\nextgroupplot[xlabel={(a) True spectrum}]
    \addplot+ [hist={density, bins=23}] table [y=True spectrum] {\datatable};

\nextgroupplot[xlabel={(c) Unc}]
     \addplot+ [hist={density, bins=23}] table [y=Unc] {\datatable};
     
\nextgroupplot[xlabel={(e) Heat}]
     \addplot+ [hist={density, bins=23}] table [y=Heat] {\datatable};
     
\nextgroupplot[xlabel={(g) BR}]
     \addplot+ [hist={density, bins=23}] table [y=Poly] {\datatable};
     
\nextgroupplot[xlabel={(b) Ref spectrum}]
     \addplot+ [hist={density, bins=23}] table [y=Ref spectrum] {\datatableref};
     
\nextgroupplot[xlabel={(d) Tr}]
     \addplot+ [hist={density, bins=23}] table [y=Tr] {\datatable};

\nextgroupplot[xlabel={(f) Sqrt}]
     \addplot+ [hist={density, bins=23}] table [y=Sqrt] {\datatable};
     
\nextgroupplot[xlabel={(h) Sq}]
     \addplot+ [hist={density, bins=23}] table [y=Sq] {\datatable};

\end{groupplot}
\end{tikzpicture}
	\caption{Histograms representing the empirical spectral distribution of different Laplacian matrices. Panels a) and b) show the histograms of the true Laplacian $\bbL^*$ and the reference Laplacian $\tbL$. Panel c) shows the histogram of the estimated $\hbL$ when no similarity constraint is used, and panels d), e), f), g) and h) show the histogram of $\hbL$ when the considered constraints are linear, heat kernel, square root, band-rejection, and quadratic, respectively.} \label{fig:spectrum_hist}
\end{figure*}

\section{Numerical results}\label{S:experiments}
We now present numerical experiments to gain intuition about the proposed graph-learning algorithm and to assess its performance.
We consider different test functions and compare the results achieved with popular graph-learning algorithms over a range of scenarios.
The code implementing the proposed algorithm and the experiments is available on GitHub\footnote{\url{https://github.com/reysam93/motif_nti}}. 

Upon proper selection of the test functions, the method proposed in this paper is robust to the graph scale ambiguity. Since in general this is not the case for most graph-learning algorithms, to provide a fairer comparison, we set the true GSO $\bbS^*$ and its estimate $\hbS$ to have unit Frobenius norm before computing the error.
The resulting error metric is given by
\begin{equation}
    \text{err}(\hbS, \bbS^*) = \left\|\frac{\hbS}{\|\hbS\|_F}-\frac{\bbS^*}{\|\bbS^*\|_F}\right\|_F^2.
\end{equation}
In addition, in the numerical experiments we focus on estimating the combinatorial Laplacian $\bbL$, so we solve the optimization problem in \eqref{E:split_constraint} by setting  the set of feasible GSOs to 
 $\ccalL := \{ L_{ij} \leq 0 \;\mathrm{for} \; i \neq j; \; \bbL=\bbL^\top; \;   \bbL \textbf{1} = \bb0\}$.
 While our algorithms work for any type of GSO, most of the literature focuses on learning Laplacians, so setting $\bbS=\bbL$ here facilitates the comparisons with the state of the art. 

\subsection{Proposed test functions}\label{S:test_functions}
The test functions $g$ are at the core of the similarity constraints proposed in this paper. 
Hence, before presenting the numerical results, we provide the different test functions considered in the experiments and the associated upper bounds.

\vspace{1mm}\noindent\textbf{Linear test function.}
Considering $g(x)=x$ results in the similarity constraint \eqref{E:tr_constraint}.
Since it involves the $\tr(\bbS)$, we denote it as ``Tr'' in the experiments.
This function renders the similarity constraint convex so no upper bound is required.

\vspace{1mm}\noindent\textbf{Heat kernel test function.}
Setting $g(x)=e^{-x}$ results in a convex function $c_g$ with an associated upper bound $u(\bblambda, \bblambda^{(t-1)})=\frac{1}{N}\sum_{i=1}^N\lambda_i e^{-\lambda^{(t-1)}}$.
This is denoted as ``Heat'' in the experiments.

\vspace{1mm}\noindent\textbf{Square root test function.}
Setting $g(x)=\sqrt{x}$ results in a concave function $c_g$ with an associated upper bound $u(\bblambda,\bblambda^{(t)}))=\frac{1}{2N}\sum_{i=1}^N\frac{\lambda_i}{\lambda^{(t-1)}_i}$.
This is denoted as ``Sqrt'' in the experiments.

\vspace{1mm}\noindent\textbf{Quadratic test function.}
Setting $g(x)=x^2$ results in a convex function $c_g$ with an associated upper bound $u(\bblambda, \bblambda^{(t-1)}) = -\frac{2}{N}\sum_{i=1}^N\lambda^{(t-1)}_i\lambda_i$.
This is denoted as ``Sq'' in the experiments.

\vspace{1mm}\noindent\textbf{Band-rejection test function.} Setting $g(x)=(x-1.5)^2/4$ results in a convex function $c_g$ with an associated upper bound given by $u(\bblambda, \bblambda^{(t-1)}) = \frac{1}{N}\sum_{i=1}^N(0.75-0.5\lambda^{(t-1)}_i)\lambda_i$.
This test function concentrates around small and large values of $\bblambda$, resembling a band-rejection filter.
This is denoted as ``BR'' in the experiments.

\subsection{Results on synthetic graphs}
By using synthetic data we can test the algorithms in a wider range of settings, facilitating getting insights.
In the following experiments, the graph signals $\bbX=[\bbx_1,...,\bbx_M]$ are sampled from a GMRF where the covariance matrix is given by the pseudo-inverse of the true Laplacian denoted as $(\bbL^*)^\dagger$.
The reported error corresponds to the mean error averaged across 100 realization of random graphs and graph signals.

\vspace{1mm}\noindent\textbf{Test case 1.}
The first experiment probes how the test functions in Section~\ref{S:test_functions} influence the spectrum of the estimated graphs.
We generate the target graph $\ccalG$ and the reference graph $\tilde{\ccalG}$ as two lattice graphs with 4 neighbors and  $N=200$ and $\tilde{N}=150$ nodes, respectively.
The histograms of their eigenvalues $\bblambda$ and $\tblambda$ are depicted in Fig.~\ref{fig:spectrum_hist}a and Fig.~\ref{fig:spectrum_hist}b, where we can observe that the spectra of both graphs are clearly similar.
Then, the remaining panels show the spectrum of the estimated GSOs, $\hblambda$, obtained following Algorithm~\ref{A:NTI_tight} when no similarity constraint is employed (Fig.~\ref{fig:spectrum_hist}c), as well as for the different test functions.
It can be seen that employing any of the selected similarity constraints renders the empirical distribution of $\hblambda$ closer to the ground truth than not using any constraint.
It is also worth noting that ``Heat'' and ``Sqrt'' test functions (Fig.~\ref{fig:spectrum_hist}e and Fig.~\ref{fig:spectrum_hist}f) properly capture the distribution of low-valued eigenvalues but struggle with high-valued eigenvalues, resulting in longer tails.
On the other hand, ``BR'' and ``Sq'' test functions (Fig.~\ref{fig:spectrum_hist}g and Fig.~\ref{fig:spectrum_hist}h) are better suited for capturing the shape of the distribution associated with medium and large eigenvalues, but are less precise with the smaller ones. 
This interesting behavior could help in designing specific test functions that efficiently capture the shape of the spectral distribution of the graph, a worth-looking problem that is considered as a future research direction.  

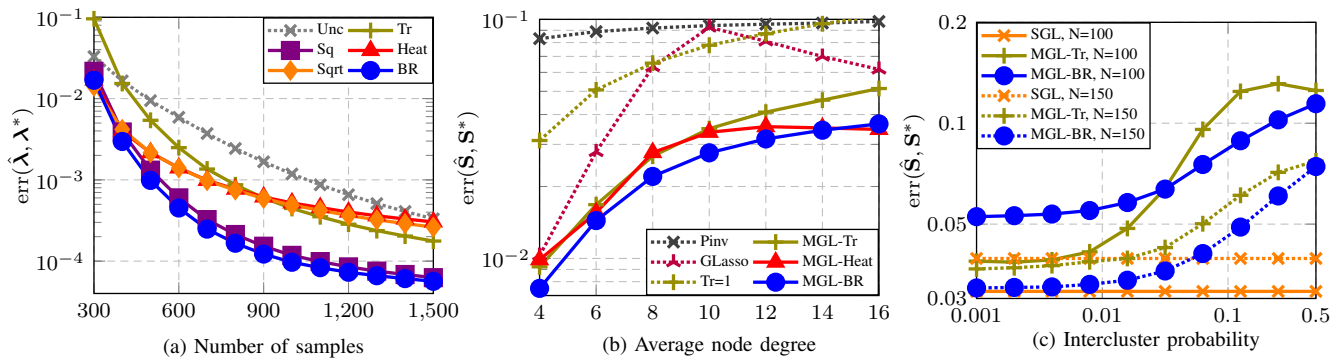
\begin{figure*}[!t]
	\centering
	\begin{subfigure}{0.32\textwidth}
		\centering
		 \begin{tikzpicture}[baseline,scale=1]

\pgfplotstableread{data/constraints.csv}\errtable

\begin{semilogyaxis}[
    xlabel={(a) Number of samples},
    xmin=300,
    xmax=1500,
    xtick={300,600,...,1500},
    ylabel={$\mathrm{err}(\hblambda,\bblambda^*)$},
    ymin=0.00004,
    ymax=0.1,
    ytick={.0001,.001,.01,.1},
    grid style=densely dashed,
    grid=major,
    legend style={
    at={(1,1)},
    anchor=north east},
    legend columns=2]
    
    \addplot[gray, mark=x, densely dotted] table [x=xaxis, y=Unc] {\errtable};
    \addplot[olive, mark=+] table [x=xaxis, y=Tr] {\errtable};
    \addplot[violet, mark=square*] table [x=xaxis, y=Sq] {\errtable};
    \addplot[red, mark=triangle*] table [x=xaxis, y=Heat] {\errtable};
    \addplot[orange, mark=diamond*] table [x=xaxis, y=Sqrt] {\errtable};
    \addplot[blue, mark=*] table [x=xaxis, y=BR] {\errtable};
    
    \legend{Unc, Tr, Sq, Heat, Sqrt, BR}
    
    
    
\end{semilogyaxis}
\end{tikzpicture}
	\end{subfigure}
	\begin{subfigure}{0.32\textwidth}
		\centering
        \begin{tikzpicture}[baseline,scale=1]

\pgfplotstableread{data/graph_density.csv}\errtable

\begin{semilogyaxis}[
    xlabel={(b) Average node degree},
    xmin=4,
    xmax=16,
    xtick={4,6,...,16},
    ylabel={$\mathrm{err}(\hbS,\bbS^*)$},
    ymin=0.007,
    ymax=0.1,
    grid style=densely dashed,
    grid=both,
    legend style={
    at={(1,0)},
    anchor=south east},
    legend cell align=left,
    transpose legend,
    legend columns=3]


    \addplot[darkgray, mark=x, densely dotted] table [x=xaxis, y=Pinv] {\errtable};
    \addplot[purple, mark=Mercedes star, densely dotted] table [x=xaxis, y=GLasso] {\errtable};
    \addplot[olive, mark=+, densely dotted] table [x=xaxis, y={MGL-Tr=1}] {\errtable};
    \addplot[olive, mark=+] table [x=xaxis, y=MGL-Tr] {\errtable};
    \addplot[red, mark=triangle*] table [x=xaxis, y=MGL-Heat] {\errtable};
    \addplot[blue, mark=*] table [x=xaxis, y=MGL-Poly] {\errtable};
    
    \legend{Pinv, GLasso, {Tr=1}, MGL-Tr, MGL-Heat, MGL-BR}

    
    
\end{semilogyaxis}
\end{tikzpicture}
	\end{subfigure}
	\begin{subfigure}{0.32\textwidth}
		\centering
		\begin{tikzpicture}[baseline,scale=1]

\pgfplotstableread{data/ref_graph_errs.csv}\errtable

\begin{loglogaxis}[
    log ticks with fixed point,
    xlabel={(c) Intercluster probability},
    xlabel style={yshift=3pt},
    xmin=10^(-3),
    xmax=.5,
    xtick={0.001,0.01,0.1,0.5},
    ylabel={$\mathrm{err}(\hbS,\bbS^*)$},
    ymin=0.03,
    ymax=0.2,
    ytick={.03, .05,.1,.2},
    grid style=densely dashed,
    grid=major,
    legend style={
    at={(0,1)},
    anchor=north west},
    legend columns=1]
    
    \addplot[orange, mark=x] table [x=xaxis, y={SGL100}] {\errtable};
    \addplot[olive, mark=+] table [x=xaxis, y={MGL-Tr100}] {\errtable};
    \addplot[blue, mark=*] table [x=xaxis, y={MGL-Poly100}] {\errtable};
    \addplot[orange, mark=x, densely dotted] table [x=xaxis, y={SGL150}] {\errtable};
    \addplot[olive, mark=+, densely dotted] table [x=xaxis, y={MGL-Tr150}] {\errtable};
    \addplot[blue, mark=*, densely dotted] table [x=xaxis, y={MGL-Poly150}] {\errtable};
    
    \legend{{SGL, N=100}, {MGL-Tr, N=100}, {MGL-BR, N=100}, {SGL, N=150}, {MGL-Tr, N=150}, {MGL-BR, N=150}}
    
    
    
    
\end{loglogaxis}
\end{tikzpicture}
	\end{subfigure}
	\caption{Mean error of the estimated GSOs when using synthetic data and different types of graphs. a) $\ccalG$ and $\tilde{\ccalG}$ are generated as lattice graphs with 4 neighbors per node; b) $\ccalG$ and $\tilde{\ccalG}$ are sampled from a small world model; c) $\ccalG$ and $\tilde{\ccalG}$ are sampled from an SBM model. All figures show error averaged over 100 realizations of graphs and signals.} \label{fig:synthetic_exps}
\end{figure*}

In addition to visually comparing the spectral distribution of the estimated graphs, Fig.~\ref{fig:synthetic_exps}a shows the error of the estimated eigenvalues as the number of signal observations $M$ increases.
The error is measured as $err(\hblambda, \bblambda^*)$, where the Frobenius norm is replaced by the $\ell_2$ norm of the vectors.
Once again, we observe that the worst performance is obtained when no similarity constraint is used (``Unc'' in the legend), clearly illustrating the benefit of accounting for the similarity of $\ccalG$ and $\tilde{\ccalG}$ based on their local structures.
Furthermore, we observe that the quadratic (``Sq'') and band-rejection (``BR'') test functions consistently outperform the linear constraint (``Tr'').
This supports our previous hypothesis that more sophisticated test functions are more capable of capturing the relationship between the reference and the sought graph.

\vspace{1mm}\noindent\textbf{Test case 2.}
We continue by evaluating the error of the estimated $\hbS$ when graphs are sampled from the small world random graph model~\cite{newman1999renormalization} as edge density increases.
True GSOs $\bbS^*$ have $N=100$ nodes while the reference graphs have $\tilde{N}=150$ nodes.
In both cases, the number of neighbors of each node increases as reflected in the x-axis of Fig.~\ref{fig:synthetic_exps}b. The edge rewiring probability is 0.1, and the number of observations is $M=1,000$.
The results illustrated in Fig.~\ref{fig:synthetic_exps}b compare the performance of our proposed approach with that of the following baselines: (i) ``Pinv'', which considers the naive solution given by the pseudo-inverse of the sample covariance matrix $\hbC$; (ii) ``GLasso'', which estimates $\hbS$ by means of the graphical Lasso algorithm~\cite{friedman2008sparse}; and (iii) ``Tr=N'', which solves problem \eqref{E:noncvx_problem} replacing the similarity constraint by the fixed constraint $\tr(\bbS)=N$ employed in \cite{dong2016learning}.
Our graph-learning algorithm based on similar motif densities is denoted as ``MGL'' followed by an additional label indicating the similarity constraint considered.
Looking at the results, we observe that the proposed MGL approach outperforms the other baselines independently of the selected similarity constraint.
Of special interest is the comparison between ``Tr=N'' and ``MGL-Tr'' since the two constraints are intimately related as discussed in Section~\ref{S:convex_relaxation}.
The results show that ``MGL-Tr'' clearly outperforms ``Tr=N'', which was expected because the first case employs information about the true value of the $\tr(\bbS)$.
Moreover, since the experiments are conducted with $\bbS=\bbL$, the value of $\tr(\bbS)$ represents the sum of the degrees across nodes, so the trace constraint can be interpreted as approximately fixing the value of $\|\bbS\|_{1}$ to its true value.

\vspace{1mm}\noindent\textbf{Test case 3.}
The last result involving synthetic graphs is portrayed in Fig.~\ref{fig:synthetic_exps}c and its objective is twofold.
It evaluates the robustness of the similarity constraints and offers a comparison with the spectral graph learning (SGL) algorithm in \cite{kumar2019structured}.
The SGL algorithm shows state-of-the-art performance when dealing with graphs with multiple connected components.
In this experiment, the true $\bbS^*$ is drawn from a stochastic block model (SBM)~\cite{newman2018networks} with $K=5$ communities.
The edge probability is $p=0.3$ for nodes within the same community and $q=0$ for nodes of different communities.
That is, the SBM graphs have 5 separate connected components, a setting for which the SGL algorithm is tailored for.
We consider the error of the estimated $\hbS$ for graphs with $100$ and $150$ nodes as indicated in the legend.
The number of samples is $M=1,000$ independently of the number of nodes.
On the other hand, the reference GSO $\tbS$ is drawn from an SBM with $150$ nodes and the same values of $p$ and $K$, but the value of $q$ increases progressively as indicated in the x-axis of the Fig.~\ref{fig:synthetic_exps}c.
Note that the error lines associated with the SGL algorithm remain constant since they do not depend on the reference graph.

A first observation from the results in Fig.~\ref{fig:synthetic_exps}c is that the MGL algorithm is surprisingly robust to the proposed perturbation on the reference graph.
The error remains below 0.1 even for values of $q$ that are comparable to the values of $p$.
Indeed, this phenomenon suggests that the similarity constraints are capturing information about the spectrum that goes beyond unveiling the number of zero eigenvalues.
Next, focusing on the graphs with 100 nodes (solid lines), the best performance is achieved by the SGL algorithm.
This was expected since the spectral constraints of SGL exactly capture the number of disconnected communities.
More illuminating are the results of graphs with 150 nodes (dashed lines), where it can be observed that the MGL outperforms the SGL algorithm for the two selected similarity constraints.
This change of behavior is caused because the error of SGL increases with the number of nodes, while the error of the MGL is decreasing as the graph grows.
We stress that this behavior is counter-intuitive because the number of samples remains constant independently of the number of nodes, and hence, a higher $N$ should carry a higher error.
Nonetheless, the rationale behind this result is as follows.
The functions $c_g(\bblambda)$ described in \eqref{E:test_function} may be interpreted as estimating the expectation of some test function $g$ \emph{across all the nodes of the graph}, and hence, as the number of nodes increases the estimation of this expectations improves.
As a result, the similarity constraint carries more information when the graph has $N=150$ nodes, compensating the additional error derived from estimating a larger number of edges, and hence resulting in a better estimate.
Finally, note that information about the number of zero eigenvalues can be incorporated into our proposed model seamlessly.

\subsection{Results on real-world graphs}
We close the numerical experiments by validating our proposed algorithm over two datasets with real-world graphs.

\vspace{1mm}\noindent\textbf{Student network dataset.}
In this experiment we consider two graphs with 32 nodes from the Ljubljana student network dataset\footnote{The original data can be found at {\scriptsize \url{http://vladowiki.fmf.uni-lj.si/doku.php?id=pajek:data:pajek:students}}}.
In these graphs, nodes represent students from the University of Ljubljana and the edges of the different networks capture different types of interactions among the students.
Because the same students (nodes) are represented across both selected networks, it is expected that the topology of the graphs will be related, allowing us to further asses the value of the method in this paper.
This dataset does not contain graph signals, which are created as a GMRF using $(\bbL^*)^\dagger$ as the covariance.
The combination of real graphs and synthetic data brings us the opportunity of evaluating the performance of the MGL algorithm on real graphs while ensuring that the observed signals comply with the assumed model.

\begin{figure}[!t]
	\centering
	\pgfplotsset{width=.36\textwidth}
	\begin{tikzpicture}[baseline,scale=1]

\pgfplotstableread{data/students.csv}\errtable

\begin{semilogyaxis}[
    log ticks with fixed point,
    xlabel={Number of samples},
    xmin=100,
    xmax=400,
    xtick={100,150,...,400},
    ylabel={$\mathrm{err}(\hbS,\bbS^*)$},
    ymin=0.027,
    ymax=0.13,
    ytick={.03,.04,.06,.1,.2},
    grid style=densely dashed,
    grid=both,
    legend style={
    at={(1,1)},
    anchor=north east},
    legend columns=1]
    
    \addplot[purple, mark=Mercedes star, densely dotted] table [x=xaxis, y={GLasso}] {\errtable};
    \addplot[olive, mark=+, densely dotted] table [x=xaxis, y={Tr=N}] {\errtable};
    \addplot[orange, mark=x, densely dotted] table [x=xaxis, y={SGL}] {\errtable};
    \addplot[gray, mark=star, densely dotted] table [x=xaxis, y={Unc}] {\errtable};
    \addplot[olive, mark=x] table [x=xaxis, y={MGL-Tr}] {\errtable};
    \addplot[blue, mark=*] table [x=xaxis, y={MGL-Poly}] {\errtable};
    
    \legend{GLasso, Tr=N, SGL, Unc, MGL-Tr, MGL-BR}

    
    
\end{semilogyaxis}
\end{tikzpicture}
	\caption{Error of the estimated GSO with the true and the reference graphs obtained from the Ljubljana student network dataset.
	Signals are sampled from a GMRF distribution and  the reported error is the average over 100 realizations.} \label{fig:student_network}
\end{figure}
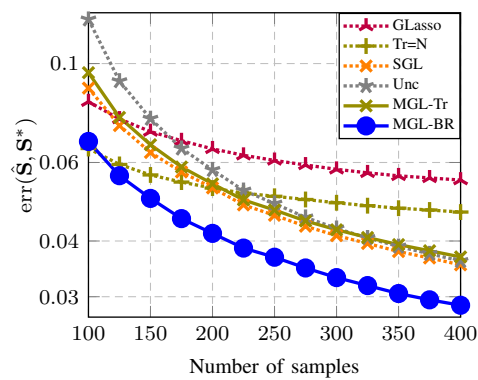

The results are depicted in Fig.~\ref{fig:student_network}, where we can observe the error of the estimated graph $\hbL$ as the number of samples increases (represented in the x-axis).
It can be seen that the MGL based on the band-rejection test function (``MGL-BR'') consistently outperforms the other alternatives.
We also note that, for the first values of the number of samples, using the fixed constraint $\tr(\bbS)=N$ renders a smaller error than using the graph similarity constraint based on the linear test function (``MGL-Tr'').
This contrast with the behavior previously observed can be explained because the number of nodes is small ($N=32$), and hence, as commented in Test case 3, the benefit of the similarity constraints is more limited.
Nonetheless, as the number of samples starts increasing the performance of ``MGL-Tr'' quickly surpasses that of ``Tr=N''.
We also observe that, for the largest values of $M$, the errors of ``SGL'', ``Unc'', and ``MGL-Tr'', seem to converge to the same value.
We recall that the ``Unc'' model is a particular implementation of the Laplacian estimation proposed in~\cite{egilmez2017graph}.

\vspace{1mm}\noindent\textbf{Senate votes dataset.}
Lastly, we consider a dataset containing the roll-call votes of the U.S. Senate~\cite{lewis2019voteview}.
As done in \cite{navarro2020joint}, we represent the congresses as networks with $50$ nodes (one per U.S. state) that encode the ideological representation of each state. Signals $\bbx_1,...,\bbx_M$ correspond to the votes on different laws and proposals. When voting on a proposal (say the $m$-th one), we codify the vote of each senator as $1$ for a yea, $-1$ for a nay, and $0$ for abstention. We then obtain the value of $\bbx_m$ for the $i$-th node as the sum of the votes of the two senators representing the $i$-th U.S. state, and repeat this process for $i=1,...,50$. 
The resultant graph signals are categorical and, thus, do not follow the assumption of being sampled from a GMRF. 
As a result, this experiment will help to illustrate that the MGL algorithm may be employed even when the observed graph signals do not follow a Gaussian distribution.

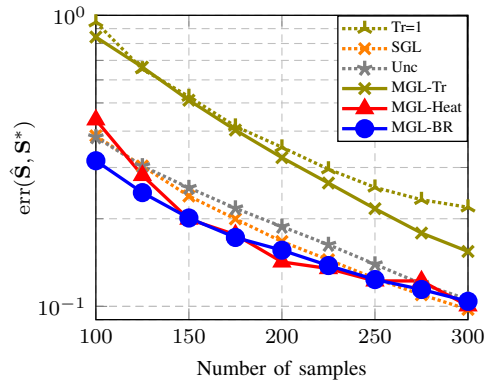
\begin{figure}[!t]
	\centering
	\pgfplotsset{width=.36\textwidth}
    \begin{tikzpicture}[baseline,scale=1]

\pgfplotstableread{data/senate.csv}\errtable

\begin{semilogyaxis}[
    xlabel={Number of samples},
    xmin=100,
    xmax=300,
    xtick={100,150,...,300},
    ylabel={$\mathrm{err}(\hbS,\bbS^*)$},
    ymin=0.09,
    ymax=1,
    grid style=densely dashed,
    grid=both,
    legend style={
    at={(1,1)},
    anchor=north east},
    legend columns=1]
    
    \addplot[olive, mark=Mercedes star, densely dotted] table [x=xaxis, y={MGL-Tr=1}] {\errtable};
    \addplot[orange, mark=x, densely dotted] table [x=xaxis, y={SGL}] {\errtable};
    \addplot[gray, mark=star, densely dotted] table [x=xaxis, y={Unc}] {\errtable};
    \addplot[olive, mark=x] table [x=xaxis, y={MGL-Tr}] {\errtable};
    \addplot[red, mark=triangle*] table [x=xaxis, y={MGL-Heat}] {\errtable};
    \addplot[blue, mark=*] table [x=xaxis, y={MGL-Poly}] {\errtable};
    
    \legend{Tr=1, SGL, Unc, MGL-Tr, MGL-Heat, MGL-BR}

    
    
\end{semilogyaxis}
\end{tikzpicture}
	\caption{Error of the estimated graph Laplacian with data obtained from the roll-call votes of the U.S. congress dataset.
	Signals are randomly sampled from the available signals and the error is the average over 100 realizations.} \label{fig:senate_network}
\end{figure}

For the setup at hand, we set the graph corresponding to the 114th congress (years 2015 and 2016) as the known reference graph $\tilde{\ccalG}$, and our goal is to estimate $\ccalG$, the graph corresponding to the 115th congress (years 2017 and 2018).
We have access to 499 and 591 observed signals for each of the graphs.
Since there are no evident ground-truth graphs, we consider as the true underlying graphs those inferred using the unconstrained solution of problem \eqref{E:noncvx_problem} when all the signals are available.
The error of the estimated $\hbL$ is reported in Fig.~\ref{fig:senate_network}, where the x-axis denotes the $M$ observed signals considered.
For low values of $M$, the ``MGL-BR'' and ``MGL-Heat'' outperform the alternatives, even though ``Unc'' is the algorithm used to generate the ground-truth graph.
Moreover, in additional experiments, we observe that considering the median error instead of the mean, the heat test function outperforms the band-rejection test function.
Recalling that the heat test function learns small eigenvalues better than the larger ones, the superior performance of the heat test function suggests that in these networks the small eigenvalues play a more fundamental role than in previous settings.
On the other hand, as $M$ increases, the error of the different models converges towards the same value, except for ``Tr=N'' and ``MGL-Tr'', showing that the trace-based constraints struggle to capture the topological properties of this graph.

\begin{figure*}[!t]
	\centering
	\begin{subfigure}{0.24\textwidth}
		\centering
		\includegraphics[width=1\textwidth]{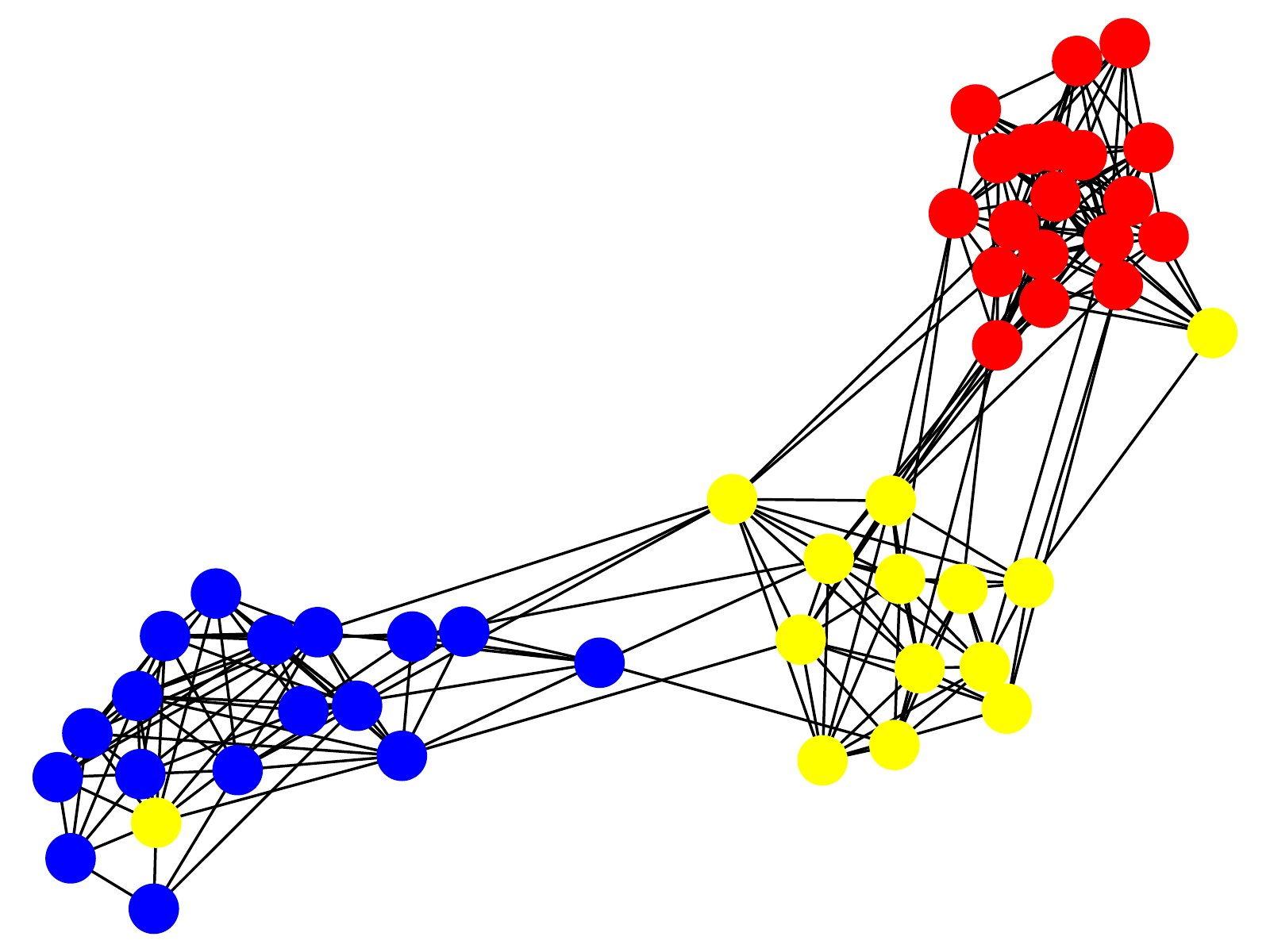}
		\caption{Ground truth.}
		\label{sfig:ground_truth}
	\end{subfigure}
	\begin{subfigure}{0.24\textwidth}
		\centering
		\includegraphics[width=1\textwidth]{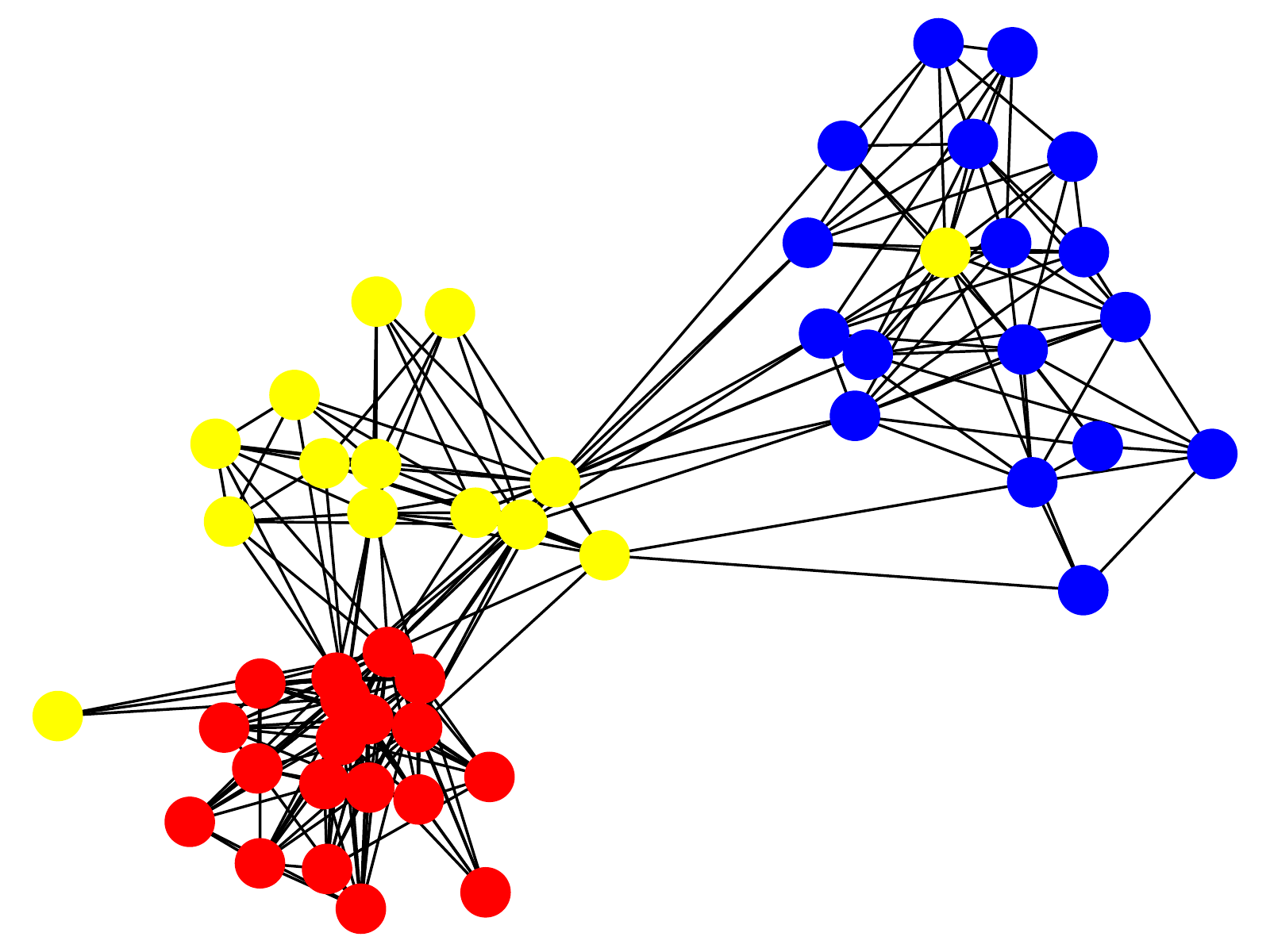}
		\caption{MGL-BR, 100 samples.}
		\label{sfig:mgl-br_100}
	\end{subfigure}
	\begin{subfigure}{0.24\textwidth}
		\centering			\includegraphics[width=1\textwidth]{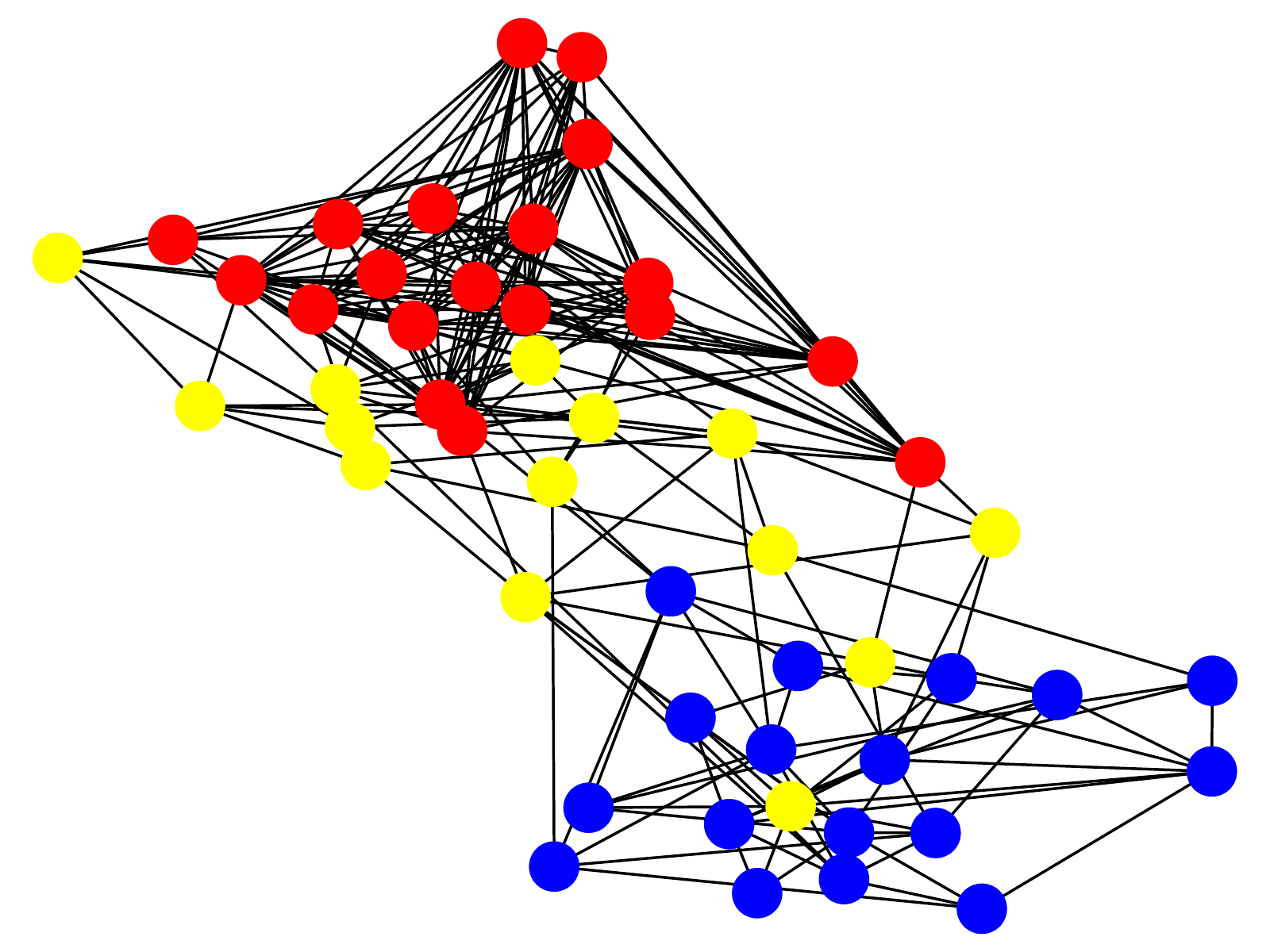}
		\caption{SGL, 100 samples.}
		\label{sfig:sgl_100}
	\end{subfigure}
	\centering
	\begin{subfigure}{0.24\textwidth}
		\centering
		\includegraphics[width=1\textwidth]{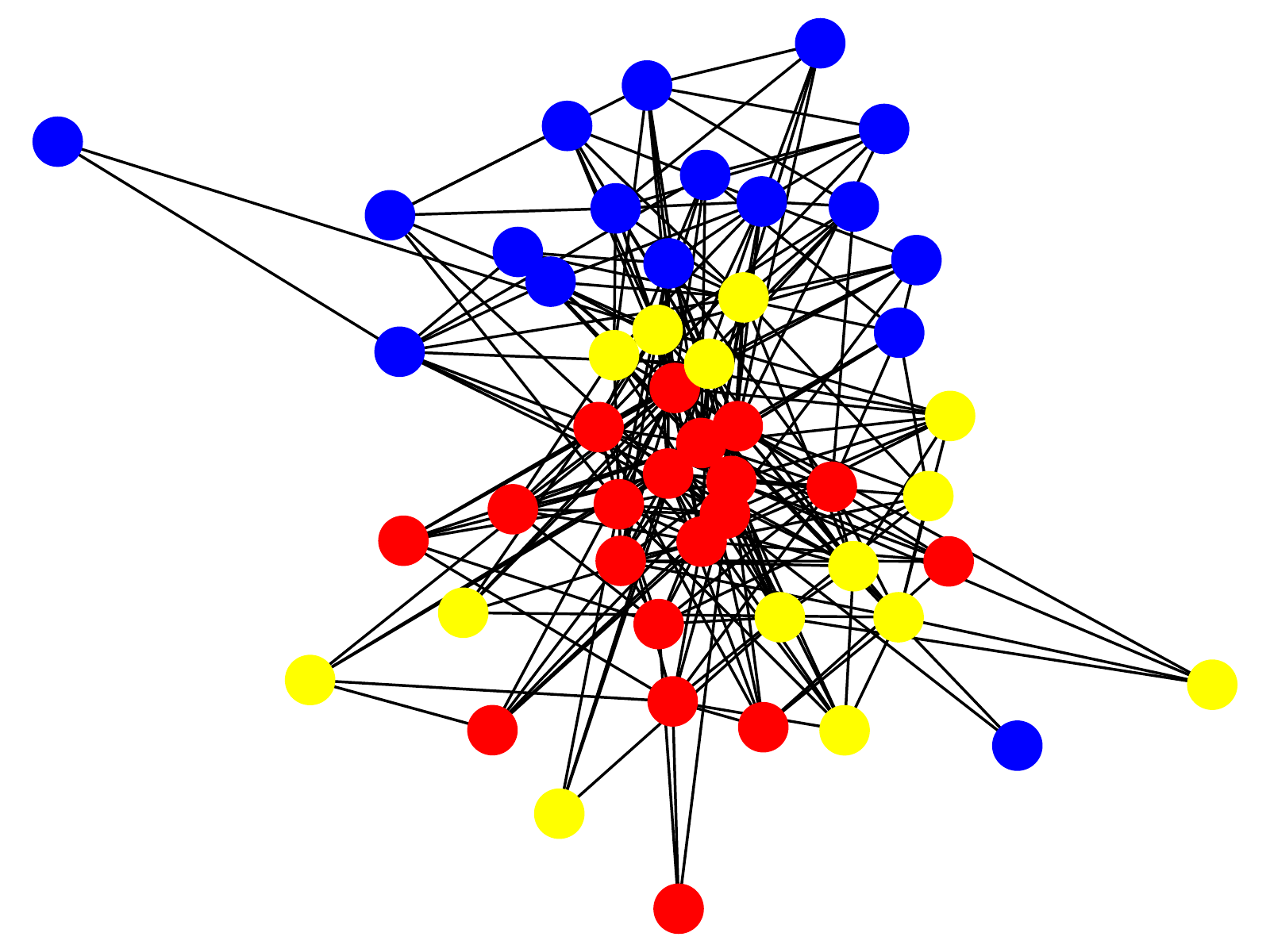}
	    \caption{Unc, 100 samples.}
	    \label{sfig:unc_100}
	\end{subfigure}
	\begin{subfigure}{0.24\textwidth}
		\centering
		\includegraphics[width=1\textwidth]{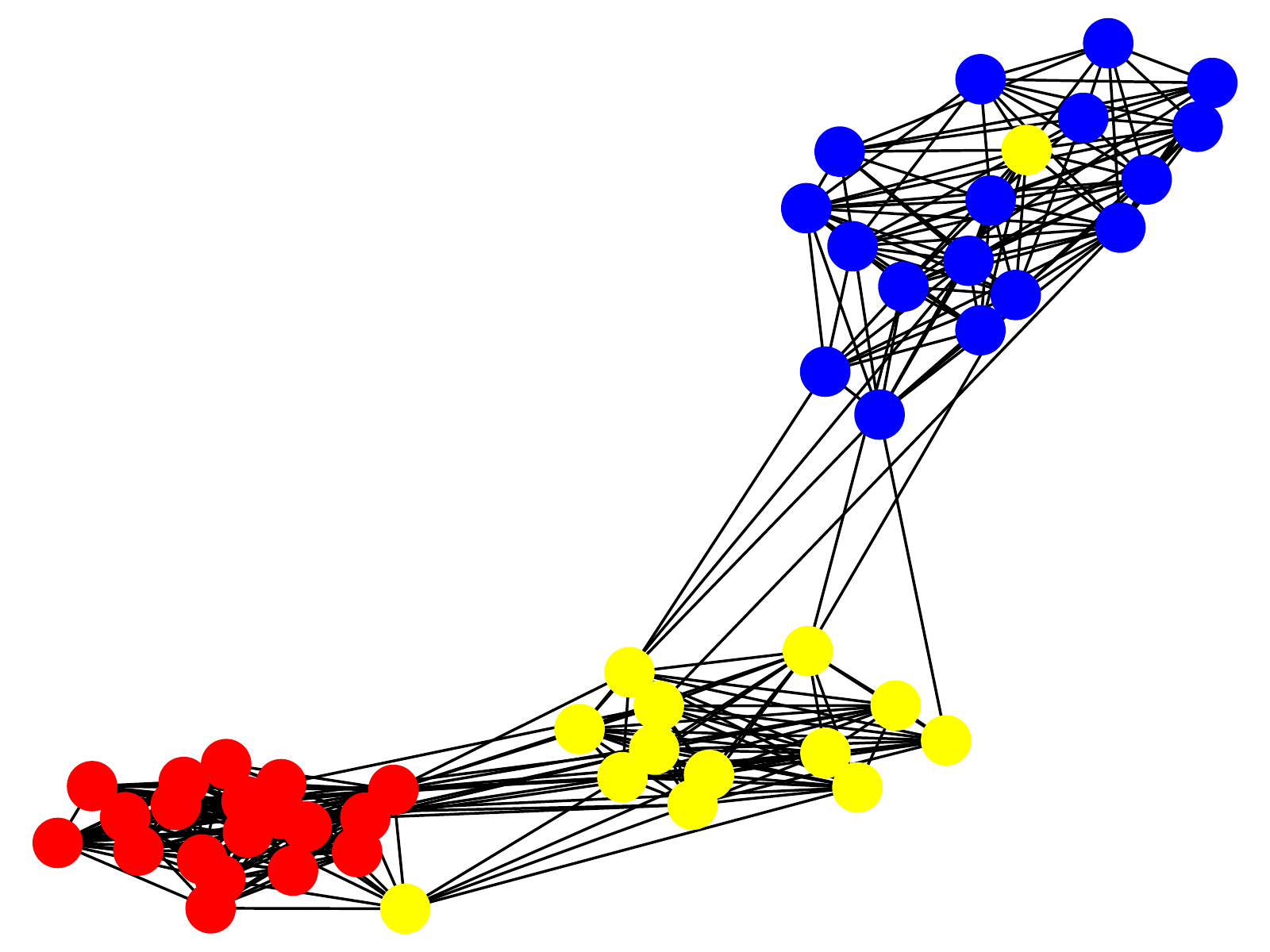}
	    \caption{MGL-Heat, 150 samples.}
	    \label{sfig:mgl-heat_150}
	\end{subfigure}
	\begin{subfigure}{0.24\textwidth}
		\centering
		\includegraphics[width=1\textwidth]{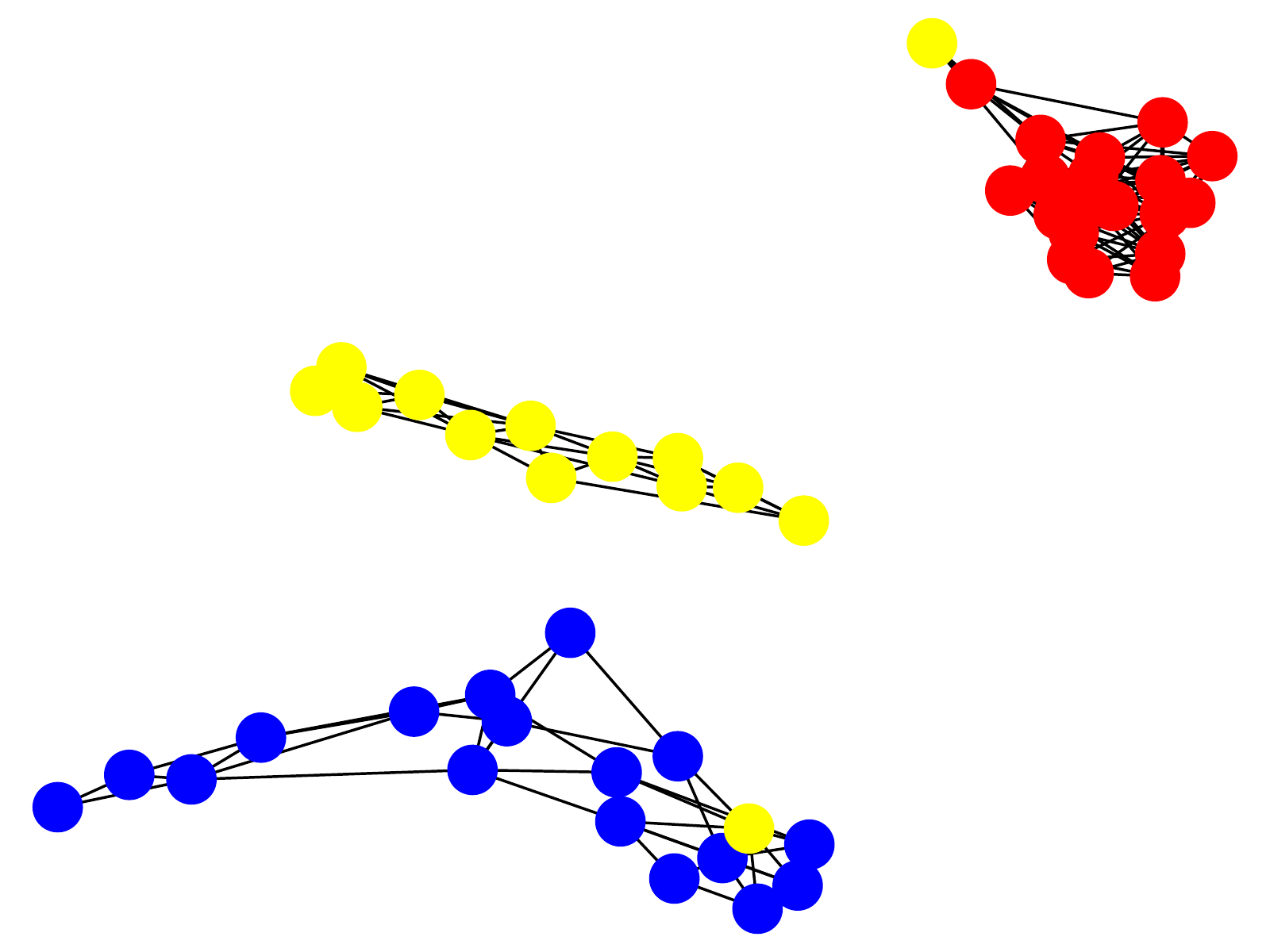}
	    \caption{MGL-BR, 150 samples.}
	    \label{sfig:mgl-br_150}
	\end{subfigure}
    \begin{subfigure}{0.24\textwidth}
		\centering
		\includegraphics[width=1\textwidth]{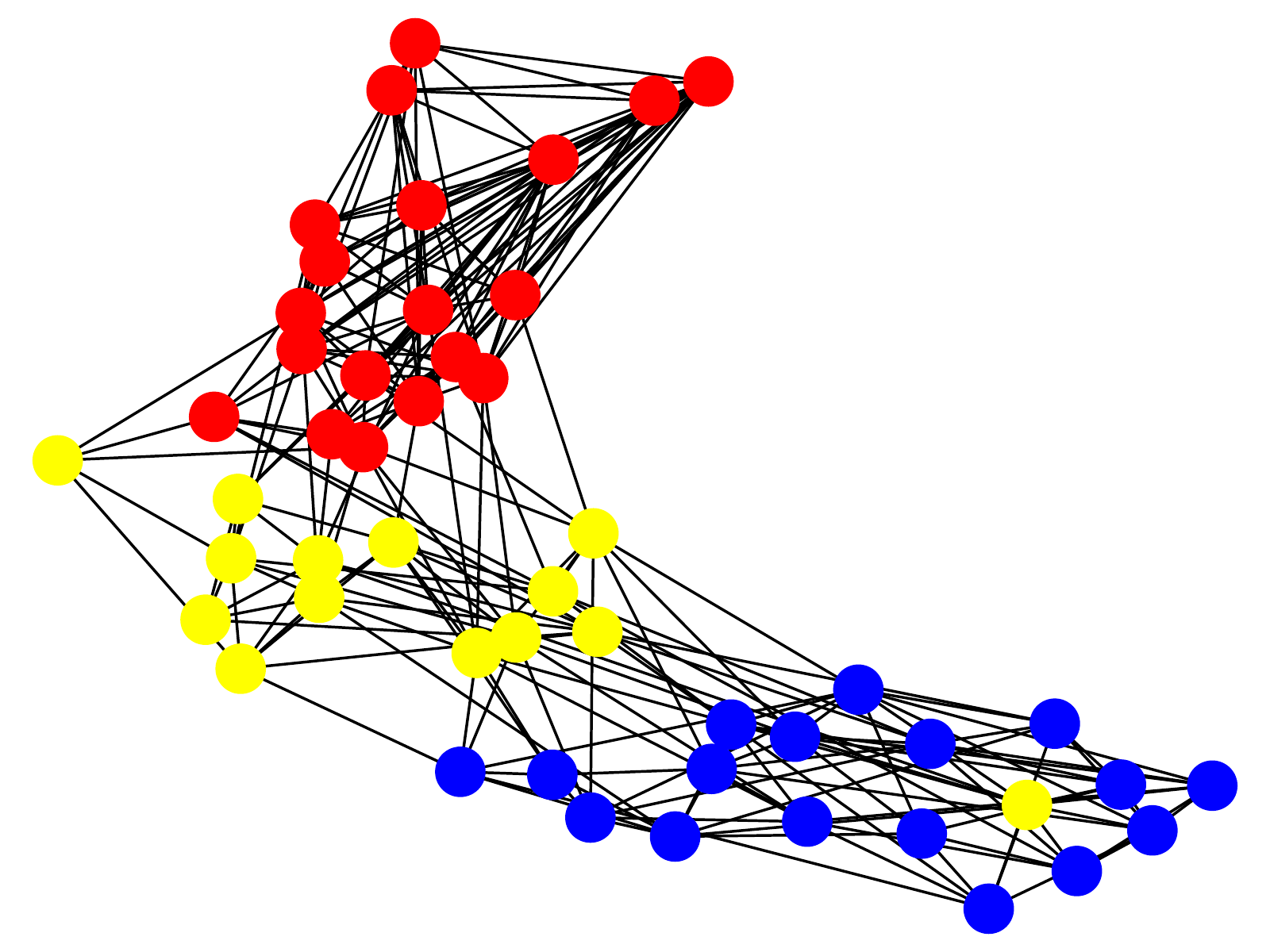}
	    \caption{SGL, 150 samples.}
	    \label{sfig:sgl_150}
	\end{subfigure}
	\begin{subfigure}{0.24\textwidth}
		\centering
		\includegraphics[width=1\textwidth]{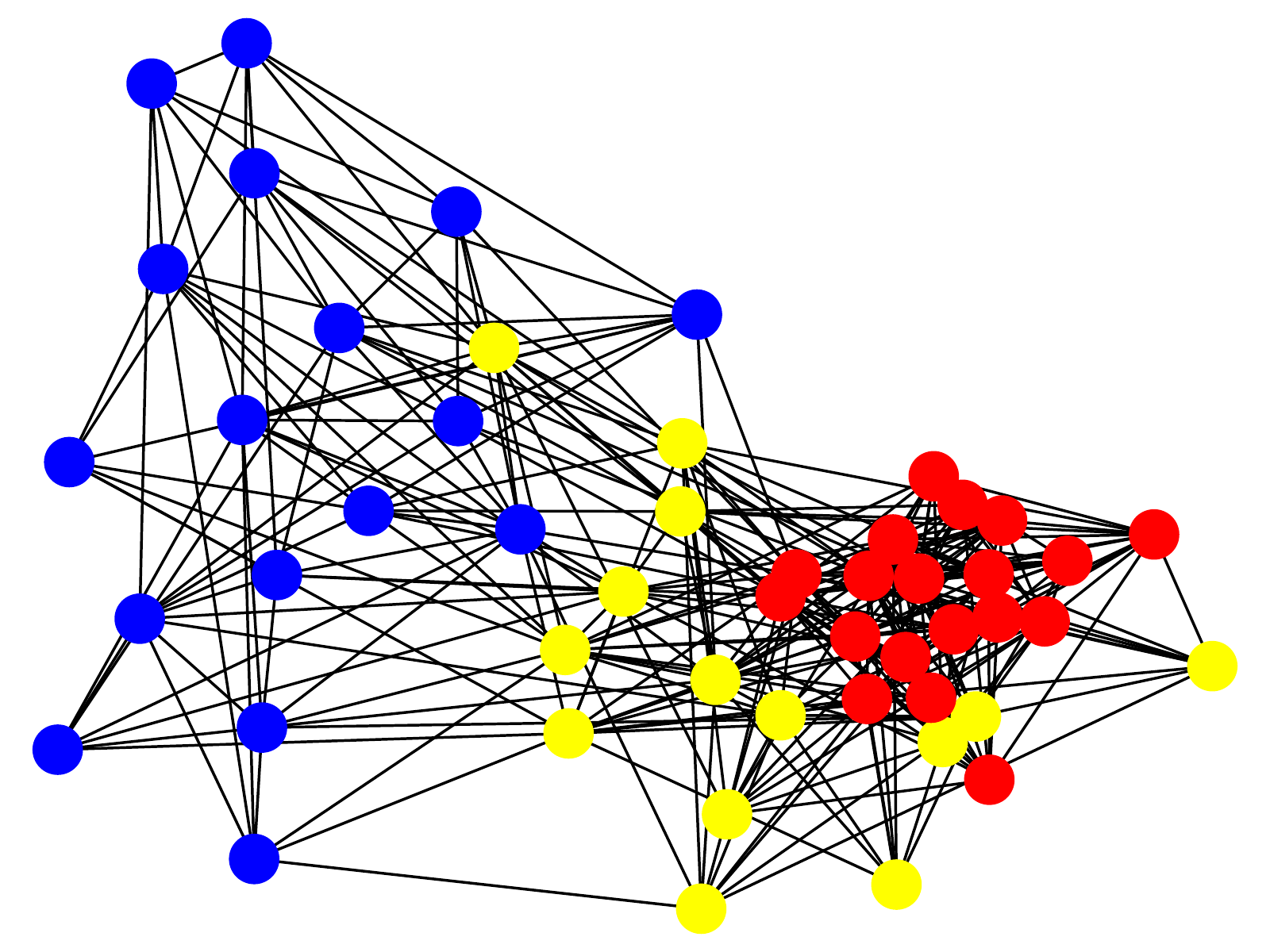}
	    \caption{Unc, 150 samples.}
	    \label{sfig:unc_150}
	\end{subfigure}
	\caption{
	Representation of the ground-truth graph and different estimations of the senate network for the 115th congress. Graphs are plotted using a spring layout and only the 250 edges with the largest weight are included. a) Shows the true graph, b), c) and d) show the estimates obtained with ``MGL-BR'', ``SGL'' and ``Unc'' algorithms when 100 samples are available, while e), f), g) and h) show the estimates obtained with ``MGL-Heat'', ``MGL-BR'',  ``SGL'', and ``Unc'' when 150 samples are available.
     }\label{fig:recovered_graphs}
\end{figure*}

To further assess the performance of the proposed algorithm, \cref{fig:recovered_graphs} compares the topology of the true graph (\cref{sfig:ground_truth}) with the estimates obtained with different graph learning algorithms.
The nodes of the graphs are colored according to the ideological representation of each estate with red, blue and yellow nodes corresponding to estates with two Republican senators, two Democratic senators, and senators from different parties, respectively.
When only 100 samples are employed, we observe that using the ``MGL-BR'' algorithm (\cref{sfig:mgl-br_100}) is the only alternative that preserves the cluster structure of the original graph.
In contrast, all the nodes are mixed in a single cluster in the ``Unc'' solution (\cref{sfig:unc_100}), and the ``SGL'' solution (\cref{sfig:sgl_100}) has the yellow nodes mixed with the red and the blue nodes.
When 150 signals are employed, it can be seen that the estimate ``MGL-Heat'' (\cref{sfig:mgl-heat_150}) is the alternative that keeps the nodes in different clusters further away while maintaining a single connected component.
Finally, the ``MGL-BR'' estimate with 150 samples (\cref{sfig:mgl-br_150}) segregates the nodes in three connected component, which may result useful in node classification or clustering tasks.

\section{Conclusion}\label{S:conclusion}
In this paper, we faced the relevant problem of learning the topology of a graph from a set of GMRF nodal observations.
The novel framework proposed herein departs from the maximum likelihood estimator of the sought graph $\ccalG$ and then exploits the assumption that the motif density of a known graph $\tilde{\ccalG}$ is similar to that of $\ccalG$.
Indeed, comparing the density of motifs of two graphs is a non-trivial combinatorial task that we addressed by leveraging a relation between the distribution of the spectra of both graphs.
More precisely, we showed that, when two graphs have similar motif densities, evaluating a continuous test function over their respective empirical distribution of eigenvalues renders a similar value.
This observation was exploited as a constraint in an optimization problem.
The resulting similarity constraints were non-convex for most test functions, so we also developed a convex relaxation by proposing an efficient iterative algorithm capable of handling any differentiable convex or concave test function.
The proposed algorithm blends techniques from MM algorithms and alternating optimization, it is guaranteed to converge to a stationary point, and its computational complexity is cubic in the number of nodes.
Finally, we evaluated the proposed algorithm through different numerical experiments involving synthetic and real-world data, where we assessed the influence of several test functions and showed that the proposed algorithm outperforms other popular alternatives.

\appendices
\section{Proof of Theorem~\ref{T:spectral_closeness}}\label{A:spectral_closeness}
Let $\bbS$ and $\tbS$ denote the GSOs of $\ccalG$ and $\tilde{\ccalG}$, and let $\bblambda\in\reals^N$ and $\tblambda\in\reals^{\tilde{N}}$ denote their respective eigenvalues.
From \textbf{(AS1a)}, it follows that $\lambda_i$ is contained in a bounded interval of $\reals$ for every $i$, so the spectrum of $\bbS$ has compact support.
The same holds for $\tbS$.
Denote the union of the supports of both empirical spectral densities by $A$.
According to the Stone-Weierstrass theorem~\cite{weierstrass1885}, any continuous function defined over a compact domain can be approximated arbitrarily and uniformly well by polynomials.
That is to say, there is some polynomial of degree $r$, which we denote by $g_r$, such that for all $\lambda\in A$, it holds that $|g(\lambda)-g_r(\lambda)|\leq\delta_1$ for some $\delta_1\geq 0$.
Moreover, $\delta_1\to 0$ as $r\to\infty$.
One can then see that
\begin{equation}
    |c_g(\bblambda)-c_{g_r}(\bblambda)|\leq\delta_1,
\end{equation}
with the same bound holding for $\tilde{\bblambda}$.

Let $\{\alpha_r^{(k)}\}_{k=1}^K$ be an enumeration of all isomorphism classes of rooted $r$-balls whose underlying graph satisfies \textbf{(AS1a)}.
Define the function $h$ on these rooted $r$-balls so that for each $\alpha_r^{(k)}$, $h(\alpha_r^{(k)})$ yields the diagonal entry at the root of the polynomial $g_r$ applied to the GSO of $\alpha_r^{(k)}$.
Since there are only finitely many such rooted $r$-balls, the magnitude of $h$ is bounded by some constant $C\geq 0$.

Let $\rho$ be a node in $\ccalG$.
Then, if the rooted ball $V_r(\ccalG,\rho)$ is isomorphic to $\alpha_r^{(k)}$ for some $k$, we have that
\begin{equation}
    [g_r(\bbS)]_{ii}=h(\alpha_r^{(k)})=h(V_r(\ccalG,\rho)).    
\end{equation}
Since $\ccalG$ satisfies \textbf{(AS1a)}, every rooted $r$-ball $V_r(\ccalG,\rho)$ satisfies \textbf{(AS1a)}, so that we can write
\begin{equation}\label{E:funtion_f}
    c_{g_r}(\bblambda)=\frac{1}{N}\sum_{i=1}^N h(V_r(\ccalG,i))=\sum_{k=1}^K h(\alpha_r^{(k)})\tau_r(\alpha_r^{(k)},\ccalG),
\end{equation}
with a similar equality holding for $c_g(\tilde{\bblambda})$ and $\tilde{\ccalG}$.
By \textbf{(AS1b)}, we have
\begin{align}
    |c_{g_r}(\bblambda)-c_{g_r}(\tilde{\bblambda})| &\leq \sum_{k=1}^K |h(\alpha_r^{(k)})|\cdot|\tau_r(\alpha_r^{(k)},\ccalG)-\tau_r(\alpha_r^{(k)},\tilde{\ccalG})| \nonumber\\
    &\leq \min\{K,\max\{N,\tilde{N}\}\}\cdot C\epsilon.
\end{align}

We conclude the proof via a simple application of the triangle inequality.
\begin{align}
    |c_g(\bblambda)-c_g(\tilde{\bblambda})| &\leq |c_g(\bblambda)-c_{g_r}(\bblambda)|+ |c_{g_r}(\bblambda)-c_{g_r}(\tilde{\bblambda})| \nonumber\\
    &\qquad+ |c_{g_r}(\tilde{\bblambda})-c_g(\tilde{\bblambda})| \nonumber\\
    &\leq 2\delta_1+\min\{K,\max\{N,\tilde{N}\}\}\cdot C\epsilon =: \delta.
\end{align}

\ifCLASSOPTIONcaptionsoff
  \newpage
\fi

\section{Efficient approximation for step 1}\label{A:step1}
We follow the procedure from \cite{kumar2019structured} to develop an efficient solution for \eqref{E:step1}.
The main difference is that we consider any suitable GSO while the cited work focus on the particular case where $\bbS$ is a combinatorial graph Laplacian $\bbL$.

We start by exploiting the symmetry of the GSO.
To that end, recall that $\bbcalS:\bbs\in\reals_+^{N(N-1)/2}\to\bbcalS\bbs\in\reals^{N \times N}$ denotes the linear operator mapping the non-negative vector $\bbs$ into the matrix $\bbS=\bbcalS\bbs$ while ensuring that the constraints in $\ccalS$ are satisfied.
Also, recall that $\|\bbS\|_1=\tr(\bbS\bbH)$, where $\bbH$ is an $N\times N$ matrix of signed ones with the sign of its entries matching the sign of the entries of $\bbS$, so we have that $\tr(\hbC\bbS)+\alpha\|\bbS\|_1=\tr(\bbK\bbS)$, where $\bbK=\hbC+\bbH$.

Then, we rewrite the problem in \eqref{E:step1} as 
\begin{alignat}{2}\label{E:lower_tri}
    \!\!&\!\ \bbs^{(t+1)} \!= \argmin_{\bbs} \
    &&\tr(\bbK\bbcalS\bbs) \!+\! \frac{\beta}{2}\|\bbcalS\bbs \!-\! \bbV^{(t)}\bbLambda^{(t)}\bbV^{(t)^\top}\|_F^2 \nonumber \\
    \!\!&\!\hspace{1.4cm} \mathrm{s.t}: && \bbs\geq0,
\end{alignat} 
where the number of optimization variables has been reduced to less than half.
Moreover, we denote as $\bbcalS^*:\bbY\in\reals^{N\times N}\to\bbcalS^*\bbY\in\reals^{N(N-1)/2}$ the adjoint linear operator of $\bbcalS$ such that $\langle\bbcalS\bbs,\bbY\rangle=\langle\bbs,\bbcalS^*\bbY\rangle$.
Then, we reformulate \eqref{E:lower_tri} as the following equivalent quadratic problem
\begin{alignat}{2}\label{E:step1_qp} 
    \!\!&\!\ \min_{\bbs\geq0} \
    && \frac{1}{2}\|\bbcalS\bbs\|_F^2-\bbz^\top\bbs,
\end{alignat}
with $\bbz=\bbcalS^*(\bbV^{(t)}\bbLambda^{(t)})(\bbV^{(t)})^\top-\beta^{-1}\bbK)$.
Although the problem in \eqref{E:step1_qp} is strictly convex, the non-negativity constraint prevents us from obtaining a closed-form solution.
To circumvent this issue, we replace the objective function of \eqref{E:step1_qp} with an upper bound centered at $\bbs^{(t)}$, resulting in the optimization
\begin{alignat}{2}\label{E:step1_major} 
    \!\!&\!\ \min_{\bbs\geq0} \
    && \frac{1}{2}\bbs^\top\bbs-\bbs^\top\left(\bbs^{(t)}-\frac{1}{\|\bbcalS\|_2^2}\nabla f(\bbs^{(t)})\right).
\end{alignat} 
The term $\nabla f(\bbs^{(t)})=\bbcalS^*(\bbcalS\bbs^{(t)})-\bbz$ denotes the gradient of the objective function in \eqref{E:step1_qp} and $\|\bbcalS\|_2^2$ denotes the operator norm given by $\|\bbcalS\|_2^2=\sup_{\|\bbx\|=1}\|\bbcalS\bbx\|_F^2$.

Finally, the closed-form solution from the KKT optimality conditions of \eqref{E:step1_major} is given by
\begin{equation}\label{E:obj_approx_step1}
    \bbs^{(t+1)}=\left(\bbs^{(t)}-\frac{1}{\|\bbcalS\|_2^2}\nabla f(\bbs^{(t)}) \right)^+,
\end{equation}
which is the update for the first step provided in \eqref{E:step1_simplified}.

\bibliographystyle{IEEEtran}
\bibliography{main.bib}

\end{document}